\documentclass[galaxies,article,accept,moreauthors,pdftex]{Definitions/mdpi} 

\usepackage{graphicx}% Include figure files
\usepackage{subfigure}
\usepackage{mathtools}

\firstpage{1} 
\pubvolume{xx}
\issuenum{1}
\articlenumber{5}
\pubyear{2020}
\copyrightyear{2020}
\history{Received: 27 July 2020; Accepted: 7 September 2020; Published: date}
\updates{yes} % If there is an update available, un-comment this line

\setitemize{parsep=6pt,itemsep=0pt,leftmargin=*,labelsep=5.5mm}
\setenumerate{parsep=6pt,itemsep=0pt,leftmargin=*,labelsep=5.5mm}
\setlist[description]{itemsep=0mm}  
 
\usepackage{subfigure}
\makeatletter
\renewcommand{\@thesubfigure}{\normalsize(\textbf{\alph{subfigure}})}
\makeatother

\usepackage{environ}
\NewEnviron{myequation}{%
\begin{equation}
\scalebox{0.9}{$\BODY$}
\end{equation}}

\Title{A Relativistic Orbit Model for Temporal Properties of~AGN}

\Author{Prerna Rana $^{\dagger}$\orcidA{} and A. Mangalam *$^{,\dagger}$\orcidB{}}
\AuthorNames{Prerna Rana and A. Mangalam}

\address[1]{
Indian Institute of Astrophysics, Sarjapur Road, 2nd Block Koramangala, Bangalore~560034, India; prernarana@iiap.res.in}

\corres{\hangafter=1 \hangindent=1.05em \hspace{-0.82em}Correspondence: mangalam@iiap.res.in} %

\firstnote{\hangafter=1 \hangindent=1.05em \hspace{-0.82em}These authors contributed equally to this work.
}

\abstract{We present a unified model for X-ray quasi-periodic oscillations (QPOs) seen in Narrow-line Seyfert 1 (NLSy1) galaxies, $\gamma$-ray and optical band QPOs that are seen in Blazars. The origin of these QPOs is attributed to the plasma motion in corona or jets of these AGN. In the case of X-ray QPOs, we applied the general relativistic precession model for the two simultaneous QPOs seen in NLSy1 1H 0707-945 and deduce orbital parameters, such the radius of the emission region, and spin parameter $a$ for a circular orbit; we obtained the Carter's constant $Q$, $a$, and the radius in the case of a spherical orbit solution. In other cases where only one X-ray QPO is seen, we localized the orbital parameters for NLSy1 galaxies REJ 1034+396, 2XMM J123103.2+110648, MS 2254.9-3712, Mrk 766, and MCG-06-30-15. By applying the lighthouse model, we found that a kinematic origin of the jet based $\gamma$-ray and optical QPOs, in a relativistic MHD framework, is possible. Based on the inbuilt Hamiltonian formulation with a power-law distribution in the orbital energy of the plasma consisting of only circular or spherical trajectories, we show that the resulting Fourier power spectral density (PSD) has a break corresponding to the energy at ISCO. Further, we derive connection formulae between the slopes in the PSD and that of the energy distribution. Overall, given the preliminary but promising results of these relativistic orbit models to match the observed QPO frequencies and PSD at diverse scales in the inner corona and the jet, it motivates us to build detailed models, including a transfer function for the energy spectrum in the corona and relativistic MHD jet models for plasma flow and its polarization properties.}
\begin{document}
%%%%%%%%%%%%%%%%%%%%%%%%%%%%%%%%%%%%%%%%%%
%%%%%%%%%%%%%%%%%%%%%%%%%%%%%%%%%%%%%%%%%%
%\setcounter{section}{-1} %% Remove this when starting to work on the template.
%\section{How to Use this Template}
\section{Introduction}
\label{intro}
Active galactic nuclei (AGN), at the center of most galaxies, are known to be powered by black holes of masses $M_{\bullet}=10^{5}-10^{9}M_{\odot}$ \cite{Rees1984,BlanfordRees1992,Antonucci1993}. These systems are believed to be the scaled-up version of black hole X-ray binaries (BHXRB), possessing the same physical process of accretion \cite{McHardy2010}. The riveting evidence of this conjecture is the similarity between the X-ray variability in AGN and BHXRB \cite{McHardyetal2006,McHardy2010}. However, a~complete understanding of the physical processes of accretion and the jet emission in AGN requires the variability analysis in various wavelength bands, from optical to $\gamma$ ray.

 The detection of quasi-periodic oscillations (QPOs) in X-ray light curves is an important breakthrough in the study of accretion processes in BHXRB \cite{Remillard2006,BelloniStella2014}. There have been many detections of low-frequency \mbox{($\nu < 30$ Hz)} and high-frequency ($\nu > 30$ Hz) QPOs in BHXRB in the Milky Way and nearby galaxies, whereas the number of significant QPOs detected in AGN is small compared to those in BHXRB. Various claims of the detection of QPOs have been made in different classes of AGN, over the last decade, with timescales ranging from a few tens of minutes to hours in X-rays, days and also years in optical and $\gamma$ ray light curves~ \cite{Gierlinski2008,Linetal2013,Alstonetal2015,Ackermann2015,Sandrinelli2014,Sandrinelli2016a,Sandrinelli2016b,
 Sandrinelli2017,Sandrinelli2018,Guptaetal2009,Grahametal2015,Kingetal2013,Fanetal2014,Smith2018,Valtonen2016,
 Britzen2018,Dey2018,Valtonen2019,Dey2019,Komossa2020}. 
 
\textit{X-ray Power spectral density (PSD) shape}: The X-ray variability is a key diagnostic for understanding the physical processes in the innermost regions of the accretion flow. The similarity in the behavior of X-ray variability in AGN and BHXRB is an important aspect of the AGN-BHXRB connection. The PSD of both BHXRB and AGN are known to show red noise, which decreases steeply at high frequencies (small timescales) following a power law, $P(\nu) \propto \nu^{-\alpha}$, where $\alpha \sim2$ typically \cite{McHardy2004,Papadakis2010,MangalamWiita1993}. At lower frequencies, below a characteristic frequency, called the break frequency ($\nu_{b}$), the PSD flattens ($\alpha\sim1$) \cite{McHardy2004,Papadakis2010,MangalamWiita1993}. Such~a characteristic PSD shape is well defined by a bending power-law model \cite{McHardy2004} and found in various types of AGN with $\nu_b$ ranging from $\sim$10$^{-6}$-10$^{-4}$ Hz \cite{Papadakis2010,MartinVaughan2012}. This break frequency is expected to approximately scale as an inverse of the black hole mass. However, the bending power-law shape of the PSD shape in BHXRB is known to be associated only with the soft spectral state \cite{Cuietal1997}. Hence, the understanding of such a characteristic shape of the PSD is fundamental for probing the inner regions close to the black~hole.

\textit{X-ray QPOs}: The QPOs detected so far in the X-ray light curves of AGN are seen to be mostly associated with the Narrow-Line Seyfert 1 (NLSy1) galaxies, which are identified by the narrow width of their broad H$\beta$ emission line with FWHM$< 2000$ kms$^{-1}$, strong FeII lines, and weak forbidden lines~ \cite{OsterbrockPogge1985,Goodrich1989}. NLSy1 galaxies are also known to show rapid X-ray variability and near Eddington accretion rates \cite{Komossa2008}. The first detection of a significant QPO in an X-ray light curve was reported in RE J1034$+$396 with timescale $\sim3730$ s using the XMM-Newton data \cite{Gierlinski2008}. Another significant QPO at $\sim3.8$ hour timescale was reported in an Ultrasoft Active Galactic Nucleus Candidate 2XMM J123103.2+110648 \cite{Linetal2013}. A QPO with $\sim2$ h timescale was detected in MS 2254.9-3712 \cite{Alstonetal2015}. Later, 1H 0707-495 also showed the detection of a significant QPO at $\sim3800$s and another at timescale $\sim8265$ s with relatively low significance in the X-ray light curve \cite{Panetal2016,Zhangetal2018}. A highly significant QPO of timescale $\sim6450$ s was reported in NLSy1 Mrk 766 \cite{ZhangPengetal2017}, while another QPO (but not simultaneous) with a period of $\sim4200$ s was also reported \cite{Bolleretal2001}, making these two signals be in a $\sim$ 3:2 resonance. Another significant X-ray QPO was reported in NLSy1 MCG-06-30-15 of timescale $\sim3600$ s \cite{Guptaetal2018}. Very recently, the detection of two QPOs was reported at timescales $\sim8064.5$ s and $\sim14706$ s in ESO 113-G010 \cite{Pengetal2020}. All these X-ray QPOs, discussed above, were found in XMM-Newton data (0.3--10 keV). The connection between QPOs and 3:2 twin peaks in BHXRB, ultra-luminous X-ray sources (ULXs), and AGN was shown in \cite{Zhouetal2015} as a universal scaling of these frequencies with the black mass and spin.

\textit{Optical and $\gamma$ ray QPOs}: The optical and $\gamma$ ray QPOs are also known to be discovered in a few BL Lacertae objects, also known as BL Lac. These objects are a class of AGN characterized by their large polarization, high variability, and weak emission lines \cite{Falomo2014,Padovani2017}. These objects are interpreted as systems with a relativistic jet pointing directly towards the line of sight of the observer; hence, the jet emission dominates in these systems, and the discovered QPOs are thought to be associated with the jets. The $\gamma$ ray QPOs are majorly discovered using the \textit{FERMI}-LAT observations (100MeV-300GeV). A $\gamma$ ray QPO of timescale $T \sim630$ days was reported in PKS2155-304 \cite{Sandrinelli2014}, where this timescale was found to be twice the optical period originally proposed by \cite{Zhang2014}. Later, the presence of both these QPO timescales was confirmed \cite{Sandrinelli2016a}. A QPO of timescale $2.18 \pm 0.08$ years was discovered in $\gamma$ ray light curve of PG 1553+113~ \cite{Ackermann2015}, where correlated oscillations were found in the radio and optical fluxes. Later, an optical QPO of similar timescale, $\sim$810 days, was confirmed in PG 1553$+$113 \cite{Sandrinelli2018}. Another $\gamma$ ray QPO with a timescale of a few months, T$\sim$280 days, was reported in PKS 0537-441, where an optical QPO of timescale $\sim$T/2 was also discovered \cite{Sandrinelli2016b}. A pair of optical and $\gamma$ ray QPO was also reported in BL Lac, having similar timescales of $\sim680$ days \cite{Sandrinelli2017,Sandrinelli2018}. Another QPO of period 34.5 days is observed in the $\gamma$-ray light curve of blazar PKS 2247-131 \cite{Zhouetal2018}. Recently, an optical QPO, having a temporal period of 44 days, was detected in the Kepler light curve of an NLSy1 galaxy KIC 9650712, which may or may not be a jet-based QPO \citep{Smith2018}. Another interesting case is OJ 287, which is a quasar with a quasi-periodic optical outburst emission cycle of 12 years. This prominent outburst is explained by a black hole binary model, where a secondary black hole interacts with the accretion disk of a much more massive primary black hole
\cite{Valtonen2016,Britzen2018,Dey2018,Valtonen2019,Dey2019,Komossa2020}. A comprehensive analysis of PSDs of 11 blazars was carried out recently, using the \textit{Fermi}-LAT gamma-ray 10-years-long light curves, where a QPO in PKS 2155-304 was confirmed \cite{Tarnopolski2020}.
 %The indication of both γ-ray and optical periods was confirmed by Sandrinelli et al. (2016a).
 %\vspace{-6pt}
 
\begin{figure}[H]
\centering
\includegraphics[scale=0.16]{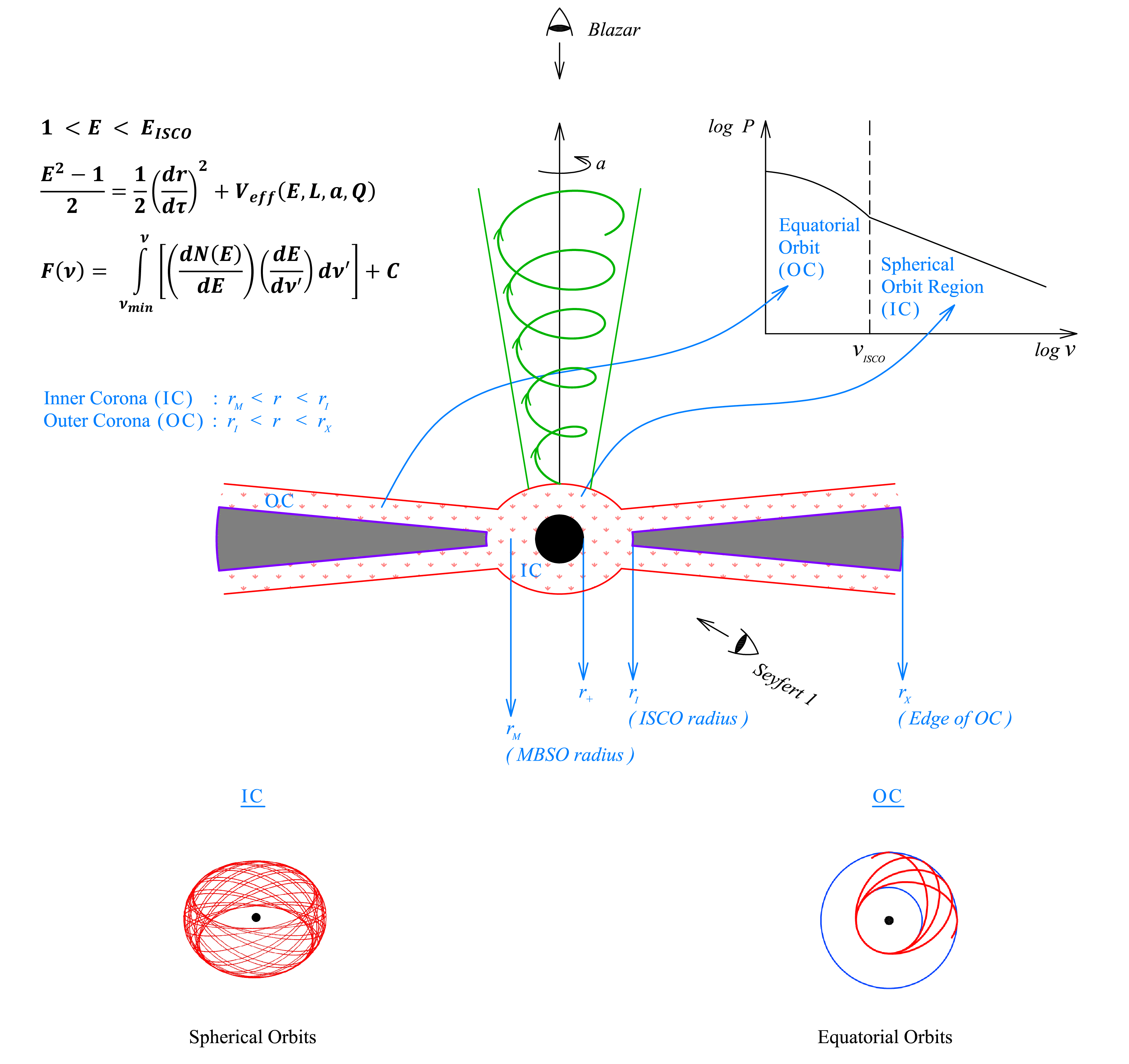}
\caption{\label{modelplot} The figure shows a unified picture of the models for X-ray, optical, and $\gamma$ ray QPOs 
 and the origin of X-ray  power spectral density (PSD) shape in AGN. The X-ray QPOs observed in NLSy1 galaxies are associated with the fundamental frequencies of the equatorial orbits in the accretion disk sandwiched by a corona region, which we call  the outer corona (OC) region, $r_I< r< r_X$; the inner corona (IC) region, $r_M< r< r_I$, is associated with the fundamental frequencies of the spherical orbits around a Kerr black hole. The optical and $\gamma$ ray QPOs in Blazars are shown as the harmonics of the timescale of a blob of matter moving along the jet. The shape of the PSD is studied using the fundamental frequency of matter which is governed by the radial effective potential, $V_{eff}(E, L, a, Q)$, providing the gravitational background responsible for the geodesic motion, in IC and OC regions to derive the energy distribution of the orbiting matter, $N(E)$, which is directly related to the observed intensity, $I(\nu)$, where $\nu$ is the temporal frequency.}
\end{figure}

In this paper, we present a model that unifies the origin of multiwavelength QPOs originating in the disk and the jet and also probes the genesis of the PSD shape of the X-ray light curve due to a corona. We study the association of X-ray QPOs discovered in NLSy1 type AGN with the relativistic circular and spherical orbits around a Kerr black hole using the generalized relativistic precession model (GRPM)~ \cite{Stella1999a,Stella1999b,RMQPO2020}. We also motivate that the non-equatorial trajectories (for example, spherical orbits), which are the consequence of axisymmetry of the Kerr spacetime, are also the viable solutions to the QPO frequencies using the GRPM \cite{RMQPO2020}. We also apply a relativistic jet model \cite{Mangalam2018,MohanMangalam2015} to study the optical and $\gamma$ ray QPO timescales in BL Lacertae objects. This model describes the simultaneous QPOs with 1:2 or 3:2 frequency ratio as the harmonics obtained in the Fourier series of the Doppler factor of the pulse profile from a blob rotating along with the jet. The Doppler factor includes the relativistic effects, such as Doppler boost, relativistic aberration, gravitational redshift, and light bending. We also present a model to describe the typical bending power-law profile of the PSD observed in AGN. Assuming the bending power-law profile of the PSD shape, we find the intrinsic profile of the energy distribution of the particles orbiting in circular and spherical trajectories in the corona around a Kerr black hole, which results in a distribution in the fundamental frequency space. The X-ray flux gets modulated at this fundamental frequency, which is a function of distance from the black hole, and consequently also a function of $E$. A unified picture of these models of multiwavelength QPOs and PSD shape is shown in Figure \ref{modelplot}, where $r_M$ is the marginally bound spherical orbit (MBSO) radius, $r_I$ is the innermost stable circular orbit (ISCO) radius, and $r_X$ is the outer disk radius.

The structure of this paper is as follows: In Section \ref{XrayQPOmodel}, we discuss the generalized relativistic precession model (GRPM) for the X-ray QPOs \cite{Stella1999a,Stella1999b,RMQPO2020}. In Section \ref{errorestimate}, we present the method for the error estimation in the parameters calculated for the case of AGN with two simultaneous X-ray QPOs, 1H 0707-495. In Section \ref{circularXrayQPO}, we discuss the association of X-ray QPO frequencies with the equatorial circular orbits using the GRPM, whereas we study the spherical orbits as the origin of X-ray QPOs using the GRPM in Section \ref{sphericalXrayQPO}. In Section \ref{Jetmodel}, we apply a basic jet model \cite{Mangalam2018,MohanMangalam2015} to study the timescales of optical and $\gamma$ ray QPOs in Blazars. We then study the genesis of the bending power-law shape of the PSD in AGN and derive the intrinsic energy distribution of the orbiting particles in Section \ref{PSDmodel}. We summarize our results in Section \ref{summary} and draw conclusions in Section \ref{Discussion}. A glossary of symbols is provided in Table \ref{glossary}. 

\begin{table}[H]
\caption{A glossary of symbols used.}
 \tablesize{\small}
 \scalebox{0.95}[0.95]{
\begin{tabular}{clcl}
\toprule
\textbf{Symbol} & \textbf{Explanation}&  \textbf{Symbol} & \textbf{Explanation} \\
\midrule
$c$ & speed of light & $\mathcal{P}_1$ & one-dimensional and normalized probability \\

$G$ & gravitational constant & &  density in parameter space\\

$M_{\bullet}$ & mass of the black hole & $L\left( a\right)$ & liklihood function for spin\\

$M_{\odot}$ & mass of the sun & $a_p$ & most probable value of spin\\

$a$ & spin of the black hole & $\chi^2_a$ & distribution function of spin\\

$Q$ & Carter's constant & $\sigma_{ai}$ & variance of spin \\

$e$& eccentricity of the orbit & $T_0$ & QPO period \\

$\mu$ & inverse-latus rectum of the orbit & $T_{F}$ & theoretical timescale for jet-based QPOs\\

$\nu$ & frequency in Hz & $r_F$ & radial footpoint of the magnetic field\\

$\bar{\nu}$ & frequency scaled by ($c^3 / G M_{\bullet}$) & $r_L$ & light cylinder radius \\

$\bar{\nu}_{\phi}$ & scaled azimuthal frequency& $\mathcal{P}_s\left( \nu\right)$ & bending power-law profile for PSD \\

$\bar{\nu}_{r}$ & scaled radial frequency & $\nu_b$ & break-frequency of PSD\\

$\bar{\nu}_{\theta}$ & scaled vertical oscillation frequency & $\alpha_l$ \& $\alpha_h$ & PSD slopes for $\nu< \nu_b$ \& $\nu> \nu_b$ \\

$r$ & radius of a circular orbit & $N\left(E \right)$ & distribution function for energy \\

$r_s$ & radius of a spherical orbit & $F\left( \nu \right)$ & distribution function for frequency\\

$p_{\theta}$ & conjugate momentum of $\theta$ coordinate & $\alpha_1$ & power-law index of $N\left(E \right)$ inside ISCO   \\

$E$ & energy per unit rest mass of a test particle &  $\alpha_2$ & power-law index of $N\left(E \right)$ outside ISCO\\

$L_z$ & z-component of the angular momentum  & $r_I$ & ISCO radius\\

& per unit rest mass of a test particle & $r_M$ & MBSO radius\\

$\tau$ & proper time & $r_X$ & outer edge of the accretion disk\\

$V_{eff}$& radial effective potential in Kerr geometry & $\bar{\nu}_I$ & scaled azimuthal frequency at ISCO \\

$P\left( \nu\right)$ & probability density in frequency space & $\bar{\nu}_M$ & scaled azimuthal frequency at MBSO\\

$\mathcal{J}$& jacobian of transformation from frequency & $\bar{\nu}_X$ & scaled azimuthal frequency at outer edge\\

& to parameter space &  &  of the accretion disk \\

$\nu_{i0}$ & observed centroid frequency of the ith QPO & $\beta_1$ & average slope of PSD for $\nu> \nu_b$ \\

$\sigma_i$ & observed standard dispersion of the ith QPO & $\beta_2$ & average slope of PSD for $\nu< \nu_b$ \\

$\mathcal{P}\left( [x]\right)$& normalized probability density in parameter & $\nu_c$ & upper cut off frequency of PSD  \\
& space & $\mathcal{P}_T$ & total integrated power of PSD \\
\bottomrule
\end{tabular}}
\label{glossary}
\end{table}

\section{Relativistic Circular and Spherical Orbits as Solutions to X-Ray QPOs}
\label{XrayQPOmodel}
 The X-ray emission from NLSy1 galaxies is believed to have originated from the inner region of the accretion disk in the context of the unification model of AGN \cite{Antonucci1993}. We apply the (G)RPM (RPM: \cite{Stella1999a,Stella1999b}; GRPM: \cite{RMQPO2020}) to study the X-ray QPOs discovered in a few cases of NLSy1 type of AGN and one Type-2 AGN candidate; see Table \ref{AGNXrayQPO}. The GRPM associates fundamental frequencies of the relativistic particle orbits in the accretion disk, close to a rotating black hole, with the QPO frequencies. Using this model, we estimate the parameters: the spin of the black hole, $a$, and radius of an equatorial circular orbit, $r$, in Kerr spacetime, where QPOs originate. We also implement the GRPM \cite{RMQPO2020} to associate the frequencies of relativistic spherical orbits (non-equatorial) with the QPO frequencies to calculate the corresponding parameters ($r_s$, $a$, $Q$), where $r_s$ is the radius of a spherical orbit and $Q$ is the Carter's constant \cite{Carter1968}, which is the fourth integral of motion in the Kerr geometry, and defined as 
\begin{equation}
\left[  p_{\theta}^2 + a^2 \cos^2 \theta  + \left(L_z^2 \csc^2 \theta -a^2 E^2 \right) \cos^2 \theta \right]=Q,  \label{Qdefntn}
\end{equation}
where $p_{\theta}$ is the conjugate momentum in $\theta$ coordinate, $L_z$ is the $z$-component of particle's angular momentum, and $E$ is its energy per unit rest mass. For the astrophysically relevant bound orbits, $Q>0$ is a valid condition for which $\theta$ obeys $0<\theta_{-} < \theta < \theta_{+} < \pi$, where $\theta_{-}+\theta_{+}=\pi / 2$, so that the orbit is symmetric with respect to the equatorial plane \cite{Carter1968,RMCQG2019}. In the equatorial plane, when $\theta=\pi /2$, $p_{\theta}$ vanishes and $\cos \theta =0$; hence from Equation \eqref{Qdefntn} we obtain $Q=0$ for the equatorial orbits.
\begin{table}[H]
\caption{A list of statistically significant QPOs detected in the X-ray band (0.3--10 keV) by the XMM-Newton in AGN along with their black masses. The lower-case letters (a to m) provide links to references given in the last column.
}
\centering
 \tablesize{\small}
 %% You can specify the fontsize here, e.g., \tablesize{\footnotesize}. If commented out \small will be used.
\begin{tabular}{cccccccc}
\toprule
\textbf{\#} & \textbf{Source}&  \textbf{Class of AGN} & \textbf{$M_{\bullet}/M_{\odot}$} & \textbf{QPO Period}& \textbf{QPO Frequency} & \textbf{References}\\
& & & \textbf{($\times 10^{6}$)} & \textbf{ks} & \textbf{$\left( \times10^{-4}\right)$Hz}& \\
\midrule
1. & RE J1034$+$396 & NLSy1 & 4 $^{a}$ & $3.73\pm0.13$  & 2.681$\pm$ 0.093  $^{\rm b}$&\cite{Zhouetal2010} $^{a}$, \cite{Gierlinski2008} $^{\rm b}$\\%MDPI: 
\midrule
2. & 2XMM J123103.2+110648  & Type-2 AGN & 0.1 $ ^{\rm c}$& $13.71$ & 0.729 $^{\rm d}$ & \cite{Hoetal2012} $^{\rm c}$, \cite{Linetal2013} $^{\rm d}$\\
\midrule
3. & MS 2254.9-3712 & NLSy1 & 4 $^{\rm e}$ & $6.667$ & $1.5$ $^{\rm f}$ & \cite{Grupeetal2004} $^{\rm e}$, \cite{Alstonetal2015} $^{\rm f}$\\
\midrule
4. & 1H 0707-495 & NLSy1 & 5.2 $^{\rm g}$ & $3.8\pm0.17$  & $2.632\pm0.118$ $^{\rm (g,h)}$ & \cite{Panetal2016} $^{\rm g}$, \cite{Zhangetal2018} $^{\rm h}$\\
& & & &$8.265\pm1.366$ & $1.21\pm 0.2$ $^{\rm h} $ & \\
\midrule
5. & Mrk 766 & NLSy1 & 4.3 $^{\rm i}$ & $6.452\pm 0.458 $ & $1.55\pm 0.11$ $^{\rm j}$ & \cite{WangLu2001} $^{\rm i}$, \cite{ZhangPengetal2017} $^{\rm j}$ \\
 & & & & $4.2$ & 2.38 $^{\rm k}$ & \cite{Bolleretal2001} $^{\rm k}$ \\
 \midrule
6. & MCG-06-30-15 & NLSy1 &3.26 $^{\rm l}$ & $3.6\pm0.229$  & 2.778$\pm$0.177 $^{\rm m}$ & \cite{Huetal2016} $^{\rm l}$, \cite{Guptaetal2018} $^{\rm m}$ \\
\bottomrule
\end{tabular}
\label{AGNXrayQPO}
\end{table}

\begin{figure}[H]
\begin{center}
\mbox{
\subfigure[]{
 \includegraphics[scale=0.3]{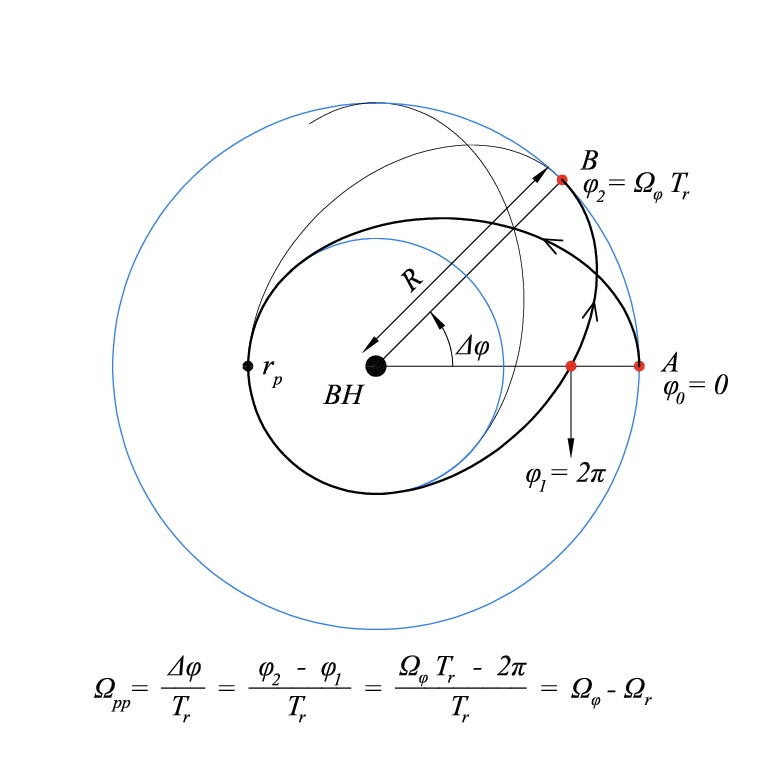}} 
 \subfigure[]{\includegraphics[scale=0.31]{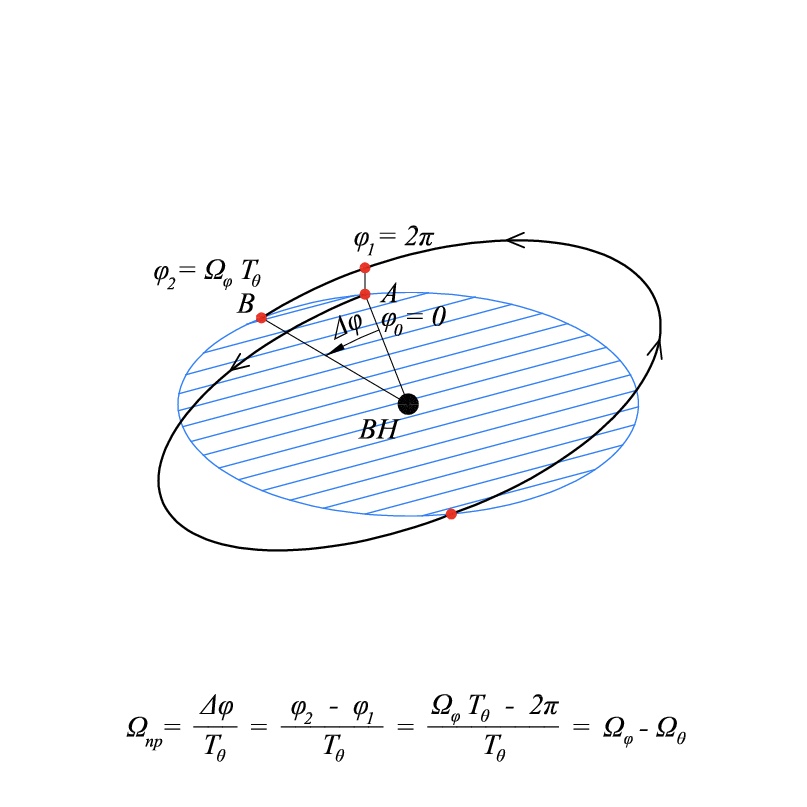}}}
 \end{center} 
 \caption{\label{precessionplots}The figure represents the generalized relativistic precession phenomenon, near a Kerr black hole (BH) at the center rotating anti-clockwise, of the non-equatorial orbits ($Q\neq0$). $\Omega_{pp}$ represents the periastron precession and $\Omega_{np}$ represents the nodal precession frequency. The particle starts from the initial point A, and follows an eccentric and non-equatorial trajectory before completing one \textbf{(a)} radial, or \textbf{(b)} vertical oscillation to reach point B, where it sweeps an extra $\Delta \phi$ azimuthal angle during one \textbf{(a)} radial, or \textbf{(b)} vertical oscillation because the azimuthal motion is faster than the radial or vertical motion. Image~courtesy: \cite{RMQPO2020}.}
 \end{figure}

The GRPM associates two simultaneous high-frequency QPOs (HFQPOs) observed in \mbox{BHXRB \cite{Remillard2006,BelloniStella2014,Stella1999a,Stella1999b,RMQPO2020}} with the fundamental frequencies: the azimuthal frequency $\left(\nu_{\phi}\right)$ and the periastron precession frequency \footnote{$pp$ stands for the periastron precession.}, $\nu_{pp}=\left(\nu_{\phi}-\nu_{r}\right)$, where $\nu_r$ is the frequency of radial oscillation. There are a few cases of BHXRB where a third low-frequency QPO (LFQPO) is also detected simultaneously to HFQPOs \cite{Mottaetal2014a,Mottaetal2014b}; in such cases, the LFQPO is associated with the nodal precession frequency \footnote{$np$ stands for the nodal precession.}, $\nu_{np}=\left(\nu_{\phi}-\nu_{\theta}\right)$, where $\nu_{\theta}$ is the frequency of vertical oscillation. See Figure \ref{precessionplots} for an illustration of the precession frequencies. In the GRPM, these frequencies are associated with the non-equatorial bound orbits ($Q\neq0$), whereas only equatorial orbits ($Q=0$) were studied in the RPM. There is no such known case in AGN where three QPOs are detected simultaneously; however, the X-ray QPO detected in Type 2 AGN 2XMM J123103.2+110648 (see Table \ref{AGNXrayQPO}) was suggested as a LFQPO because of its large rms value (25--50\%) \cite{Linetal2013}, which is the typical characterstic of LFQPOs observed in BHXRB \cite{Remillard2006}. 

Therefore, for AGN having a single QPO detection, we associate the QPO frequency with $\nu_{\phi}$, except for 2XMM J123103.2+110648 where we also analyze the $\nu_{np}$ frequency. For the cases of AGN with two simultaneous QPO detections, we use $\nu_{\phi}$ and $\nu_{pp}$ frequencies in the GRPM. In Table \ref{AGNXrayQPO}, we have summarized the cases of AGN with either one or two simultaneous QPO detections in X-rays.   

\subsection{Method for the Error Estimation}
\label{errorestimate}
Here, we describe a generic procedure \cite{RMQPO2020} which we have used to estimate errors in the orbital parameters for NLSy1 AGN with two simultaneous X-ray QPOs, 1H 0707-495; see Table \ref{AGNXrayQPO}.
\begin{enumerate}
\item We assume that the frequencies, $\nu_{1}$ and $\nu_{2}$, of QPOs are Gaussian distributed with their mean values at the centroid of observed QPO frequencies, $\nu_{10}$ and $\nu_{20}$ (with $\nu_{10}>\nu_{20}$). The joint probability density distribution of these frequencies is given by
\begin{subequations}
\begin{equation}
P\left(\nu \right) = \prod_{i=1}^{2}  P_{i}\left( \nu_{i}\right), \label{jointP}
\end{equation}
where $P_{i}\left( \nu_{i}\right)$ represents the Gaussian distribution of $i$th QPO frequency, given by
\begin{equation}
P_{i}\left( \nu_{i}\right)=\dfrac{1}{\sqrt{2\pi \sigma_{i}^2}}\exp{\left[-\dfrac{\left( \nu_{i} -\nu_{i0}\right)^2}{2\sigma_{i}^2}\right]},
\end{equation}
\end{subequations}
where $\sigma_{i}$ is the observed standard dispersion (error) of the ith QPO.
\item We find the Jacobian, $\mathcal{J}$, of the transformation from frequency to orbital parameter space using the formulae of fundamental frequencies, which is given by
\begin{equation}
 \mathcal{J}= \left[ {\begin{array}{cc}
   \frac{\partial \nu_{1}}{\partial x_1} & \frac{\partial \nu_{1}}{\partial x_2}  \\
   \frac{\partial \nu_{2}}{\partial x_1} & \frac{\partial \nu_{2}}{\partial x_2} \\
  \end{array} } \right].
\label{jacobian}
\end{equation}
where $x_1$ and $x_2$ represent the orbital parameters.
 For the equatorial circular trajectories ($Q=0$), we have \{$x_1$, $x_2$\}$=$\{$r$, $a$\}; whereas for the spherical trajectories ($Q\neq0$), we have \{$x_1$, $x_2$\}$=$\{$r_s$, $a$\}. The~Jacobian is completely expressible in an analytic form and can be easily evaluated from Equation \eqref{jacobian}, and using the frequency formulae. We utilize Equation \eqref{circfreq} for circular orbits in Section \ref{circularXrayQPO}, and Equation \eqref{sphfreq} for spherical orbits in Section \ref{sphericalXrayQPO}, to evaluate $\mathcal{J}$ (Equation \eqref{jacobian}), where $\nu_{1}=\nu_{\phi}$ and $\nu_{2}=\nu_{pp}$ according to the RPM and GRPM.

\item Next, we write the probability density distribution in the parameter space given by
\begin{equation}
P\left( [x]\right)= P\left(\nu \right)  \vert \mathcal{J} \vert, \label{Pera}
\end{equation}
where $[x]$ represents the set of parameters \{$x_1$, $x_2$\} and $\mathcal{J}$ is given by Equation \eqref{jacobian}; and \{$\nu_{1}$, $\nu_{2}$\} are substituted in terms of parameters using the analytic formulae, Equation \eqref{circfreq} for the circular orbits and Equation \eqref{sphfreq} for the spherical orbits.

\item We calculate the exact solutions for parameters by solving $\nu_{\phi}=\nu_{10}$ and $\nu_{pp}=\nu_{20}$ using Equation \eqref{circfreq} for circular trajectories \{$r_{0}$, $a_0$\}, and Equation \eqref{sphfreq} for spherical trajectories \{$r_{s0}$, $a_0$\} for fixed $Q$. We fix $M_{\bullet}$ to the previously known values. We find 1$\sigma$ errors in the parameters by taking an appropriate parameter volume around the exact solution, and generate sets of parameter combinations with resolution $\Delta x_j$ in this volume. The chosen parameter range, exact solutions, and corresponding resolutions are summarized in Tables \ref{1H07circtable} and \ref{1H07sphtable1}. We then calculate the probability density using Equation \eqref{Pera}, for all the generated parameter combinations and normalize the probability density by the normalization factor
\begin{subequations}
\begin{equation}
{\mathcal N}=\dfrac{ \sum_{k}  P\left( [x]_k\right) \Delta V_k}{V}, \ \ \Delta V_k=\prod_{j=1}^{2} \Delta x_{j , k}, \ \ V= \sum_k \Delta V_k, \label{normN}
\end{equation}
where $k$ varies from 1 to the number of total parameter combinations taken in the parameter volume; $[x]_k$ is the $k$th combination of the parameters in the parameter volume. Hence, the normalized probability density is given by
\begin{equation}
\mathcal{P}\left([x]\right)=\dfrac{P\left( [x]\right)}{{\mathcal N}}. \label{normPera}
\end{equation}
\end{subequations}
The normalization of the probability density in the parameter space, discussed above, is done because only a sub-volume in the parameter space is astrophysically allowed for bound orbits, which is discussed below. 
\item The allowed parameter combinations for the bound orbits is governed by the condition given by \cite{RMCQG2019}
\begin{equation}
\left[ \mu^3 a^2 Q \left(1+e \right)^2 +\mu^2 \left( \mu a^2 Q -x^2 -Q\right) \left(3-e \right) \left(1+e \right) +1 \right]\geq 0, \label{boundcnd}
\end{equation}
where $e$ is the eccentricity and $\mu$ is the inverse latus-rectum of the general non-equatorial trajectory. We have $e=0$ for spherical orbits; hence, we ensure that the parameters ($r_s=1/\mu$, $a$, $Q$) for spherical trajectories follow the above bound orbit condition. If any parameter combination does not obey the bound orbit condition, then $\mathcal{P}\left([x]\right)$ is taken to be zero at that point in the parameter volume.

\item For the circular orbit case, there are two parameters to estimate \{$r$, $a$\} using two QPO frequencies. For the case of spherical orbits, there are three unknown parameters \{$r_s$, $a$, $Q$\}; hence, we first take $Q = $\{1, 4, 8, 12\} for the spherical trajectory solutions, where the extrema of $\theta$ coordinate deviates away from the equatorial plane with an increase in $Q$. For each fixed value of $Q$, we find the normalized probability density distribution in the parameter space $\{x_1, x_2\}=\{r_s, a\}$ using Equation \eqref{normPera}. Later, using the calculated spin values and their errors for each fixed $Q$, we estimate the distribution of spin and the most probable spin. Using this distribution and the most probable value of the spin, we then determine the probability distribution in the $\{x_1, x_2\}=\{r_s, Q\}$ parameter space.

\item Next, we integrate the normalized probability density, $\mathcal{P}\left( [x ]\right)$, Equation \eqref{normPera}, in one dimension to obtain the profile in the other dimension. Thus, we finally obtain the one dimensional distributions \{$\mathcal{P}_{1} \left( r\right)$, $\mathcal{P}_{1} \left( a\right)$\} for circular orbits, and \{$\mathcal{P}_{1} \left( r_s\right)$, $\mathcal{P}_{1} \left( a\right)$\} for spherical orbits.

\item Finally, we fit the normalized probability density profiles in each of the parameter dimensions to find the corresponding mean values and quoted errors are obtained such that it contains a probability of 68.2\% about the peak value of the probability density. The results of these fit are given in \mbox{Tables \ref{1H07circtable} and \ref{1H07sphtable1}.}
\end{enumerate}

\subsection{Circular Orbits}
\label{circularXrayQPO}
In this section, we use the GRPM ($Q=0$) for QPOs to estimate the ($r$, $a$) parameters of the circular orbits using their fundamental frequencies, which are given by  \citep{Bardeen1972,Wilkins1972,Mottaetal2014a}
 \begin{subequations}
\begin{eqnarray}
\nu_{\phi}\left( r, a\right)=&& \frac{c^3}{2 \pi G M_{\bullet}}\frac{1}{\left( r^{3/2} + a\right)}, \ \  \bar{\nu}_{\phi}\left( r, a\right)=\frac{\nu_{\phi}}{\left( c^3 /G M_{\bullet} \right)}=\frac{1}{2\pi \left( r^{3/2} + a\right)}, \label{nuphicirc}   \\
\nu_{r}\left( r, a\right)=&& \nu_{\phi}\left( 1- \dfrac{6}{r} - \dfrac{3 a^2}{r^2} + \dfrac{8a}{r^{3/2}}\right)^{1/2}, \ \ \bar{\nu}_{r}\left( r, a\right)=\frac{\nu_{r}}{\left( c^3 /G M_{\bullet} \right)}, \label{nurcirc}  \\
\nu_{\theta}\left( r, a\right)=&& \nu_{\phi} \left( 1+ \dfrac{3 a^2}{r^2} - \dfrac{4a}{r^{3/2}}\right)^{1/2}, \ \  \bar{\nu}_{\theta}\left( r, a\right)=\frac{\nu_{\theta}}{\left( c^3 /G M_{\bullet} \right)}, \label{nuthetacirc}
\end{eqnarray}
\label{circfreq}
 \end{subequations}
where $\{ \bar{\nu}_{\phi} , \ \bar{\nu}_{r} , \ \bar{\nu}_{\theta} \}$ are the dimensionless frequencies and $M_{\bullet}$ is mass of the black hole. We use the dimensionless parameters: $r$ is scaled by $R_g=G M_{\bullet}/c^2$ and $a\equiv J /\left(G M_{\bullet}^2 /c \right)$, where $J$ is the angular momentum of the black hole. We use the convention $a>0$ for the prograde and $a<0$ for the retrograde orbits in this article.
 
\begin{figure}[H]
\centering
\mbox{
 \subfigure[]{
\includegraphics[scale=0.35]{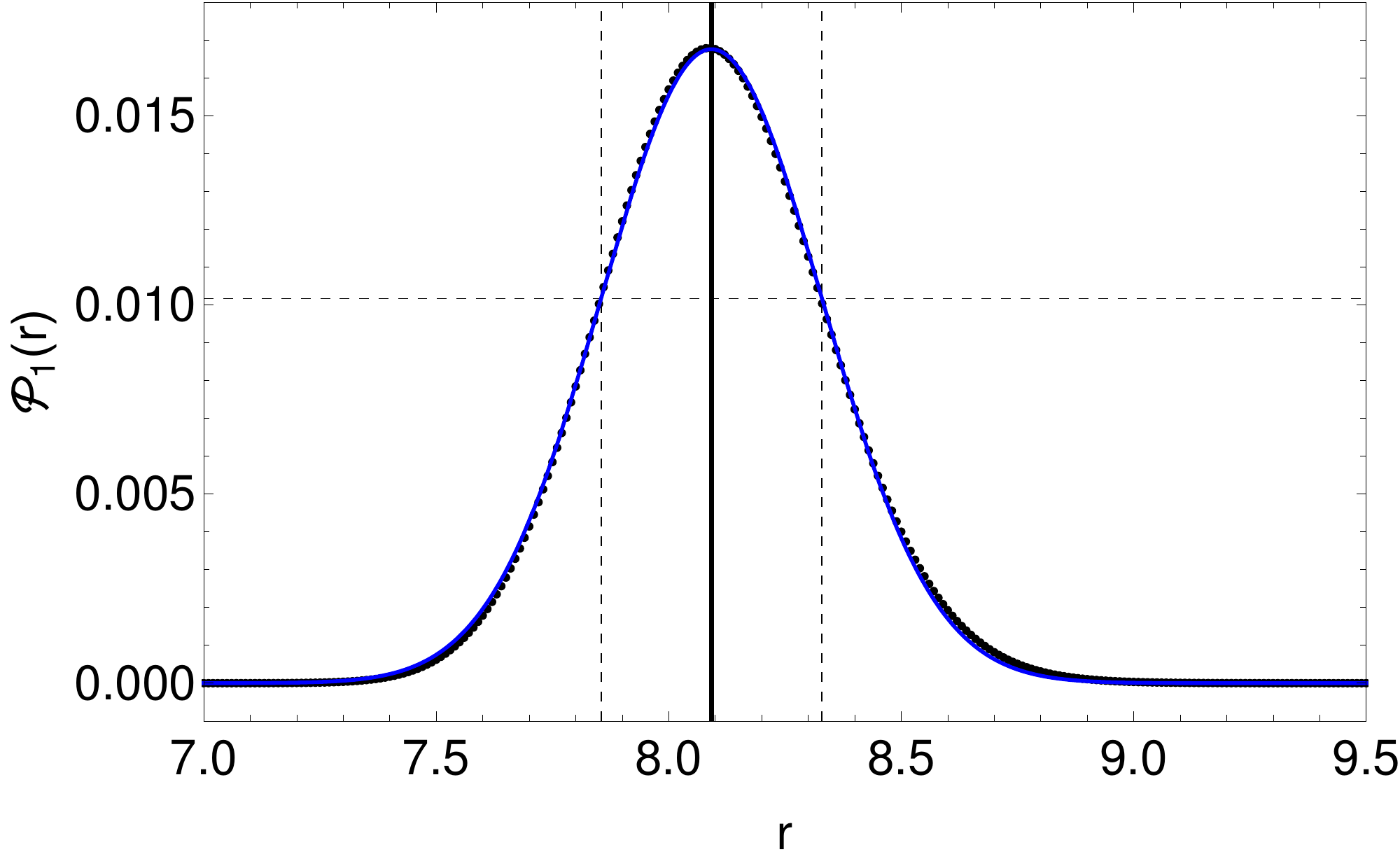}}
\hspace{0.7cm}
\subfigure[]{
\includegraphics[scale=0.42]{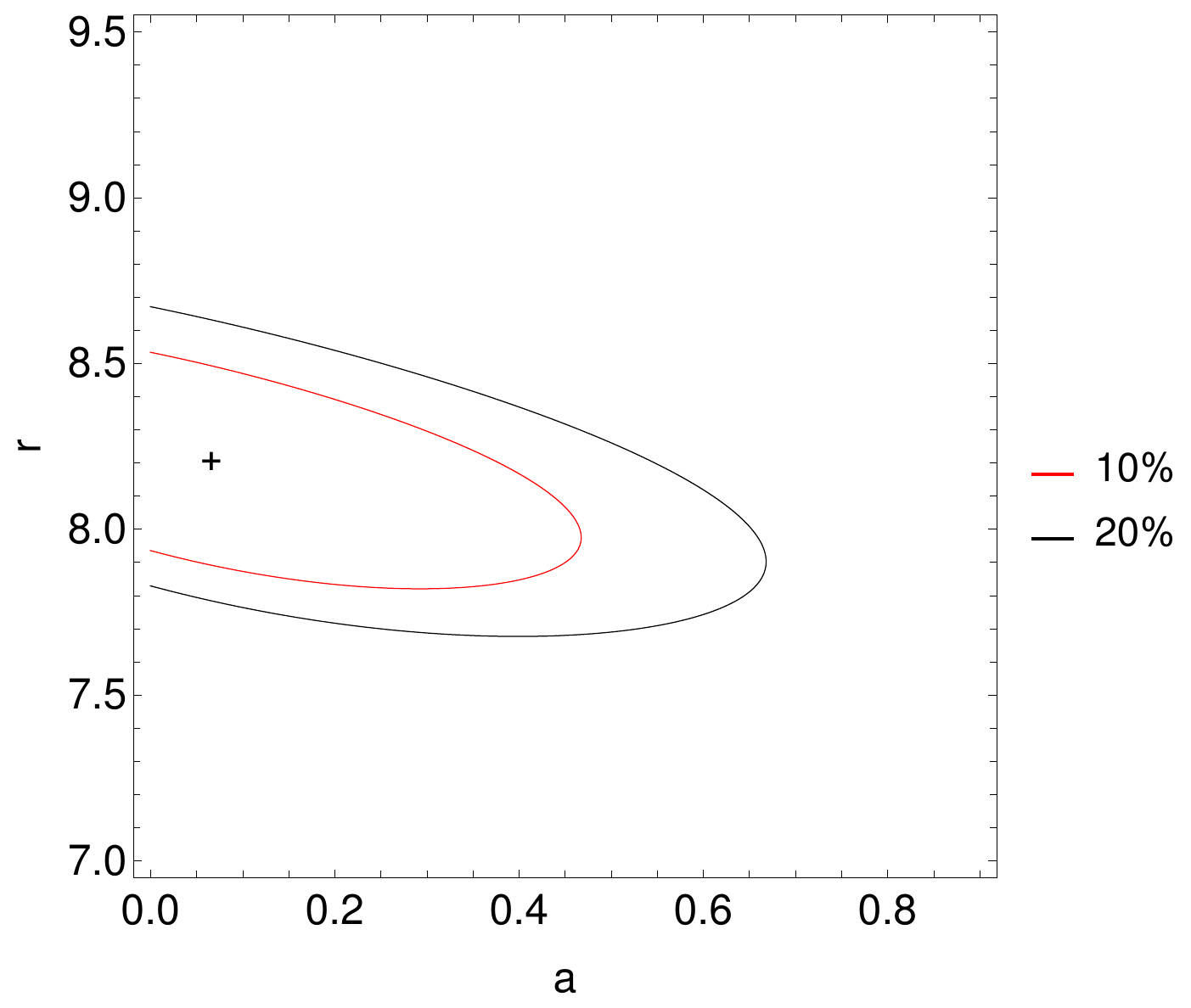}}}
\mbox{
 \subfigure[]{
\includegraphics[scale=0.42]{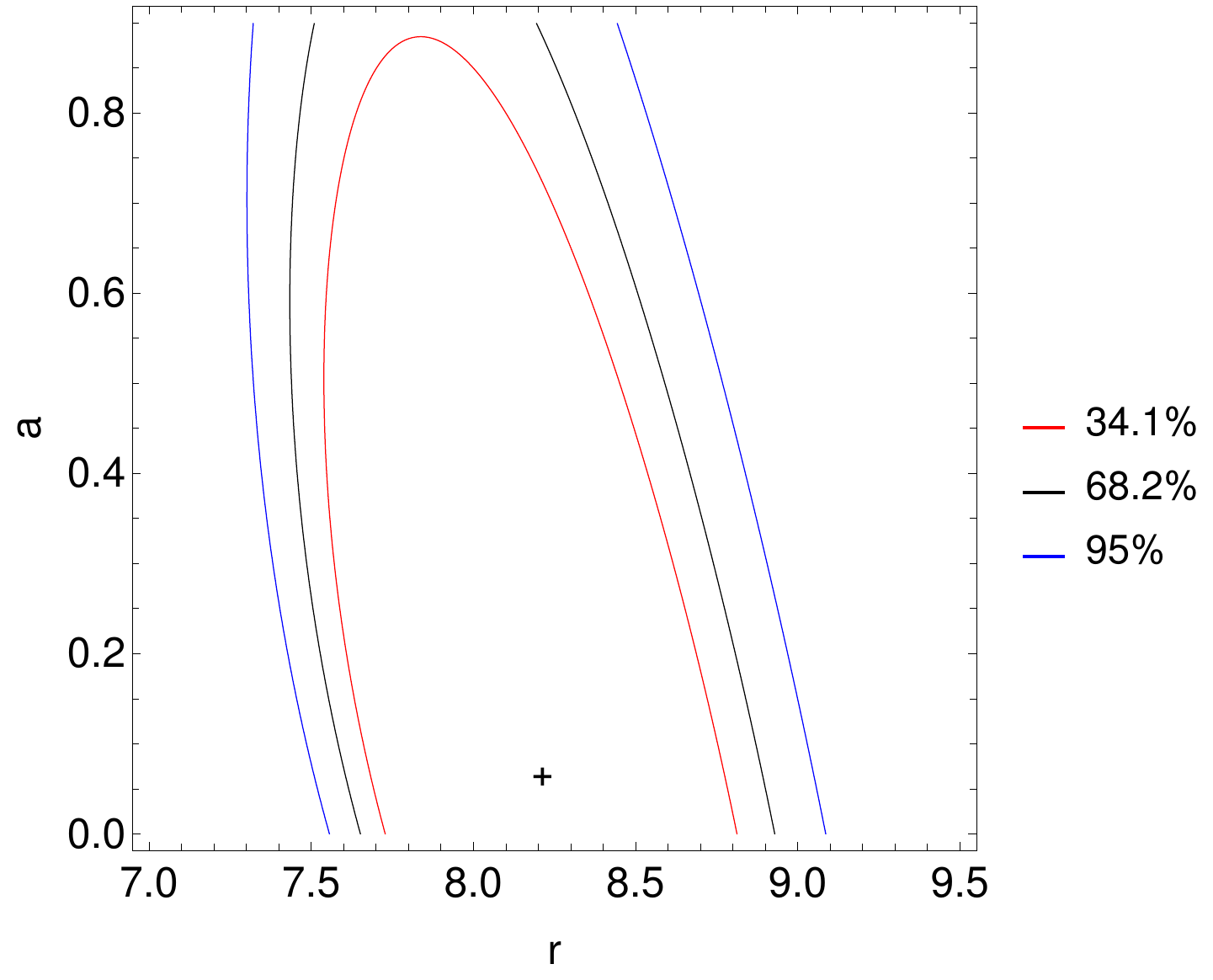}}
\hspace{0.7cm}
\subfigure[]{
\includegraphics[scale=0.35]{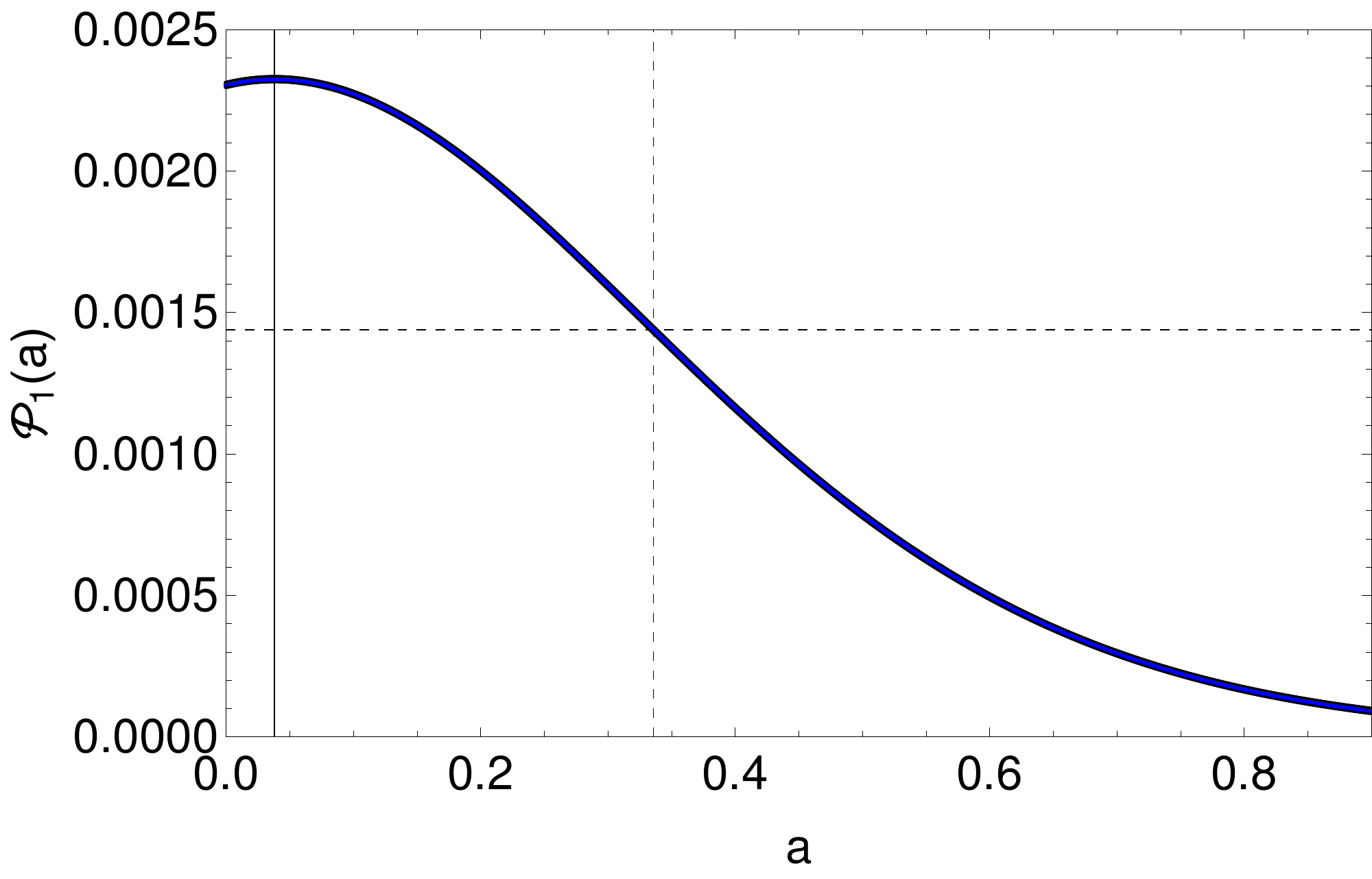}}}
\caption{\label{1H07circ}The integrated probability density profiles for 1H 0707-495 are shown in (\textbf{a}) $\mathcal{P}_1 \left( r\right)$ and (\textbf{d}) $\mathcal{P}_1 \left(a\right)$, where the dashed vertical lines enclose a region with 68.2\% probability, and the solid vertical line corresponds to the peak of the profiles. The inner probability contours of the parameter solution are shown: (\textbf{b}) in the ($a$, $r$) plane, and (\textbf{c}) the outer contours in the ($r$, $a$) plane, where the $+$ sign marks the exact solution. }
\end{figure}

 We discuss our results below:
 \begin{enumerate}
 \item We have computed the contours of $\nu_{\phi}\left( r, a\right)$, using Equation \eqref{nuphicirc}, for the QPO frequencies (given in~Table \ref{AGNXrayQPO}) of RE J1034$+$396 (blue), MS 2254.9-3712 (red), and MCG-06-30-15 (magenta), shown in the $\left( r, a\right)$ plane in Figure \ref{plotra1QPO}a. The masses of these black holes were assumed from the previous estimations (see Table \ref{AGNXrayQPO}). We see that the QPO emission originates from a very narrow region of the accretion disk, where $r\sim\left(9.4-9.9 \right)$ for RE J1034$+$396, $r\sim\left(10.4-11.4 \right)$ for MS 2254.9-3712, and $r\sim14.2 $ for~MCG-06-30-15 even though $a$ ranges from 0 to 1. This implies that the QPO emission region is very close to the black hole, and this emission region remains very narrow and nearly independent of the spin of the black hole.
 \item For the case of Mrk 766, two QPO frequencies were detected (see Table \ref{AGNXrayQPO}), but at different epochs. We have shown $\nu_{\phi}\left( r, a\right)$ contours for both these frequencies in Figure \ref{plotra1QPO}a, where \linebreak$\nu_{1}=2.38 \times 10^{-4}$Hz (orange) and $\nu_{2}=1.55\times 10^{-4}$Hz (green). The mass of the black hole was fixed to \mbox{$M_{\bullet}=4.3 \times 10^{6}M_{\odot}$ \cite{WangLu2001}}. The QPO origin range is $r\sim10$ for $\nu_{1}$ and $r\sim$ (12.6$-$14) for $\nu_{2}$, which is again found to be in a narrow range and very close to the black hole. Although these QPOs were not detected simultaneously, we tried to estimate a simultaneous solution for $\left(r, a\right)$ by equating $\nu_{\phi}=\nu_{1}$ and $\nu_{pp}=\nu_{2}$ as per GRPM. We show them as curves in the $\left(r, a\right)$ plane in Figure \ref{plotra1QPO}b, and we see that these contours do not cross each other, implying that there is no simultaneous solution for $\left(r, a\right)$.
\item For the Type-2 AGN 2XMM J123103.2+110648, the detected QPO (see Table \ref{AGNXrayQPO}) was suggested as an LFQPO type because of its large rms value \cite{Linetal2013}. If this QPO frequency is equated to the high-frequency component, $\nu_{\phi}\left(r, a\right)$, of the GRPM, we found that $r\sim200$, which is far from the black hole to emit X-rays. Hence, the GRPM predicts that this should be an LFQPO. We show the contours of the LFQPO component of the GRPM, $\nu_{np}\left(r, a\right)$, in the $\left(r, a\right)$ plane for the QPO frequency of 2XMM J123103.2+110648 in Figure \ref{plotra1QPO}c, where we fixed $M_{\bullet}=10^{5}M_{\odot}$ \cite{Hoetal2012}. We see that the emission region for this LFQPO is $r\sim\left( 6-20\right)$, for the whole range of $a$. Hence, the detected QPO of 2XMM J123103.2+110648 is an LFQPO that originated very close to the black hole.
\item For the case having two simultaneous X-ray QPOs, 1H 0707-495 (see Table \ref{AGNXrayQPO}), we first solve the equations \{$\nu_{\phi}\left(r, a \right)=\nu_{10}$, $\nu_{pp}\left(r, a \right)=\nu_{20}$\} (using Equations (\ref{nuphicirc}) and (\ref{nurcirc})), assuming $M_{\bullet}=5.2 \times 10^6 M_{\odot}$ \cite{Panetal2016}, as per GRPM to estimate the exact solution for ($r$, $a$), which is found to be ($r_0=8.214$, $a_0=0.0662$). We then apply the method, described in Section \ref{errorestimate}, to estimate the errors in the parameters ($r$, $a$) implied due to the errors of the QPO frequencies. The range of ($r$, $a$) and corresponding resolutions used for our simulations are summarized in Table \ref{1H07circtable}. Finally, we generate the probability density profiles in each parameter dimension \{$\mathcal{P}_{1} \left( r\right)$, $\mathcal{P}_{1} \left( a\right)$\}, shown in Figure \ref{1H07circ}, where we have also shown the probability contours in the parameter space. The results of the model fits to the probability density profiles are summarized in Table \ref{1H07circtable}. The errors in the parameters are quoted with respect to the exact solution ($r_0$, $a_0$), whereas the simulated \{$\mathcal{P}_{1} \left( r\right)$, $\mathcal{P}_{1} \left( a\right)$\} profiles peak at ($r=8.092$, $a=0.038$), which slightly differs from the exact solution. Hence, our analysis assuming the circular orbit frequencies as the origin of QPOs, using the GRPM, in NLSy1 1H 0707-495, suggests that it harbors a slowly rotating black hole ($a\sim0.0662$) at the center, and that the X-ray QPOs originate in the inner region of the accretion disk and very close to the black hole ($r\sim8.214$).

 \begin{table}[H]
\caption{The table summarizes results of \{$r$, $a$\} parameter solution, and corresponding errors for X-ray QPOs in NLSy1 1H 0707-495. The columns provide the range of parameter volume taken for \{$r$, $a$\}, the chosen resolution to calculate the normalized probability density at each point inside the parameter volume, the exact solutions, and the results of the model fit to the integrated profiles. The mass of the black hole is fixed to $M_{\bullet}=5.2 \times 10^6 M_{\odot}$ \cite{Panetal2016}. }
\centering
 \tablesize{\footnotesize}
  \scalebox{0.95}[0.95]{
\begin{tabular}{ccccccccc}
\toprule
\textbf{Source} & \textbf{$r$ Range} & \textbf{Resolution} & \textbf{Exact Solution}& \textbf{Model Fit} & \textbf{$a$ Range} & \textbf{Resolution} & \textbf{Exact Solution} & \textbf{Model Fit} \\
&  & \textbf{$\Delta r$} & \textbf{$r_0$} &  &  & \textbf{$\Delta  a$} & \textbf{$a_0$} & \\
\midrule
1H 0707-495 & 7--9.5 & 0.01 & 8.214 &  8.214$^{+0.116}_{-0.359}$  & 0--0.9 & 0.001 & 0.0662 &  0.0662$^{+0.2695}_{-0.0662}$  \\
&   &  & &  &  & & & \\
\bottomrule
\end{tabular}}
\label{1H07circtable}
\end{table}

 \begin{figure}[H]
\centering
\mbox{
 \subfigure[]{
\includegraphics[scale=0.3]{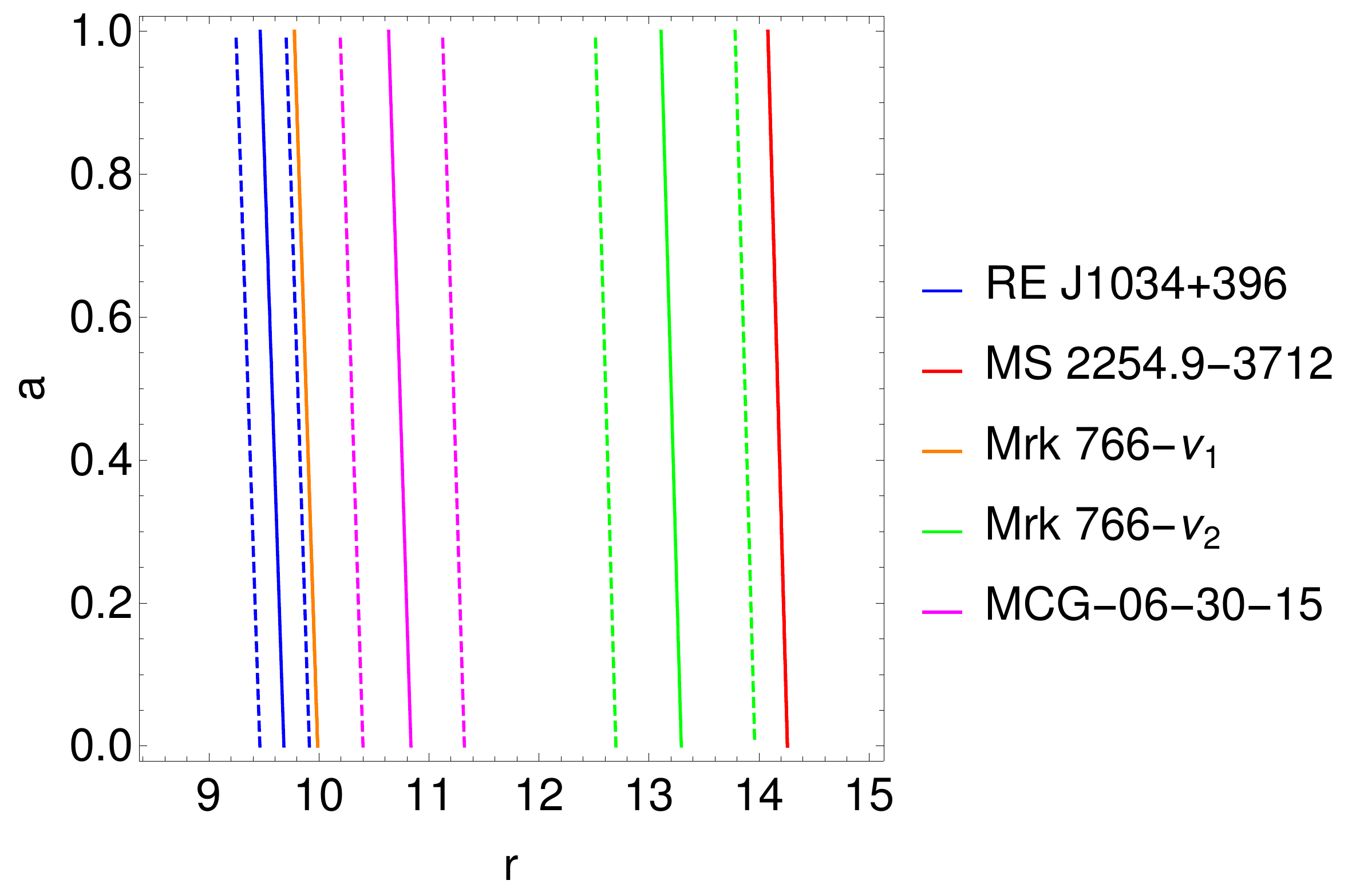}\label{raphiplot}}
\hspace{0.3cm}
\subfigure[]{
\includegraphics[scale=0.4]{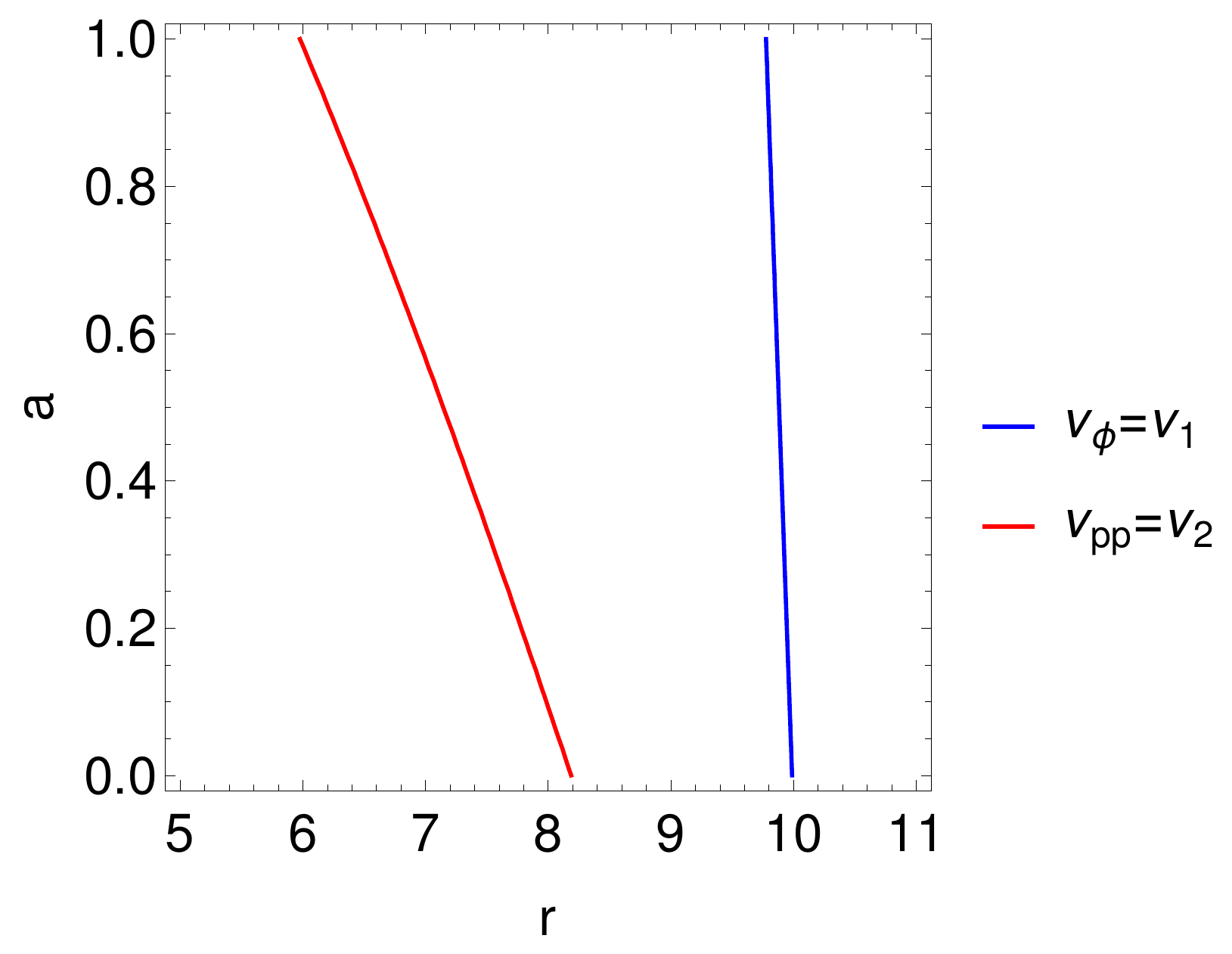}\label{Mrk766plot}}}
\mbox{
\subfigure[]{
\includegraphics[scale=0.5]{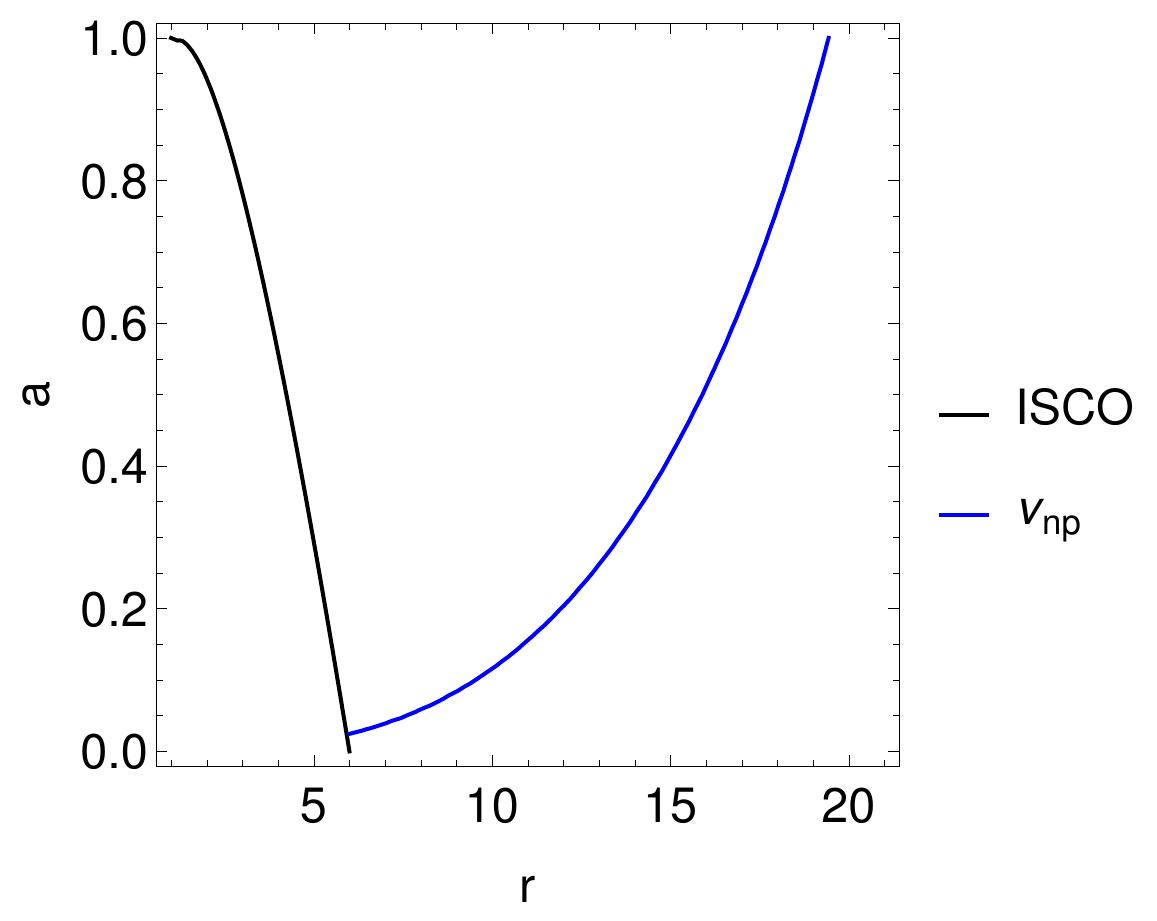}\label{plot2XMM}}}
\caption{\label{plotra1QPO}The figure shows the circular orbit frequency contours of (\textbf{a}) $\nu_{\phi}$, Equation \eqref{nuphicirc}, for the QPO frequencies of RE J1034$+$396, MS 2254.9-3712, Mrk 766, and MCG-06-30-15, given in Table \ref{AGNXrayQPO}; (\textbf{b}) $\nu_{\phi}$ and $\nu_{pp}$ contours, Equation \eqref{nurcirc}, for two QPO frequencies of Mrk 766; and (\textbf{c}) $\nu_{np}$ contour, Equation \eqref{nuthetacirc}, for the QPO frequency of 2XMM J123103.2+110648.}
\end{figure}
 \end{enumerate}

\vspace{-6pt}
\subsection{Spherical Orbits}
\vspace{-6pt}
\label{sphericalXrayQPO}
In this section, we apply the GRPM for simultaneous QPOs of 1H 0707-495 to estimate the ($r_s$, $a$, $Q$) parameters of the spherical orbits using their fundamental frequencies, which are given by \cite{Wilkins1972,RMQPO2020}
  \begin{subequations}
\begin{eqnarray}
\bar{\nu}_{\phi}\left( r_s, a, Q \right)=\frac{\left\lbrace \left[ - \dfrac{\left( 2L_z r_s -L_z r_s^2 -2r_s aE\right) }{\Delta} -  L_z \right] F\left( \frac{\pi}{2},\frac{z_{-}^{2}}{z_{+}^{2}}\right)  + L_z \cdot \Pi\left( z_{-}^2, \frac{\pi}{2},\frac{z_{-}^{2}}{z_{+}^{2}}\right)  \right\rbrace }{2 \pi \left\lbrace\left[\dfrac{\left[ E \left( a^2 r_s^2 +r_s^4 +2a^2 r_s\right) -2 L_z a r_s \right] }{\Delta}  +  a^2 z_{+}^2 E  \right] F\left( \frac{\pi}{2},\frac{z_{-}^{2}}{z_{+}^{2}}\right)  -  a^2 z_{+}^2 E \cdot K\left( \frac{\pi}{2},\frac{z_{-}^{2}}{z_{+}^{2}}\right)  \right\rbrace },
 \label{nuphisph2} \\
\bar{\nu}_r \left(r_s, a, Q \right)=\frac{\sqrt{r_s^4 \left( 1- E^2\right) + \left( 3 Q a^2 -2 x^2 r_s -2 Q r_s\right) } \cdot F\left( \frac{\pi}{2},\frac{z_{-}^{2}}{z_{+}^{2}}\right) }{2 \pi r_s \left\lbrace\left[\dfrac{\left[ E \left( a^2 r_s^2 +r_s^4 +2a^2 r_s\right) -2 L_z a r_s \right] }{\Delta}  +  a^2 z_{+}^2 E  \right] F\left( \frac{\pi}{2},\frac{z_{-}^{2}}{z_{+}^{2}}\right)  -  a^2 z_{+}^2 E \cdot K\left( \frac{\pi}{2},\frac{z_{-}^{2}}{z_{+}^{2}}\right)  \right\rbrace },
 \label{nursph2} 
\end{eqnarray}
 where $\Delta=r_s^2 +a^2 -2r_s$ and $z_{\pm}$ are given by
\begin{equation}
z_{\pm}^{2}=\frac{-P^{'}\pm \sqrt{P^{'2}-4Q^{'}}}{2}, \ \ P^{'}=\frac{-L^2_z-Q-a^2\left(1-E^2 \right) }{a^2 \left( 1- E^2\right) }, \ \
Q^{'}=\frac{Q}{a^2\left( 1- E^2\right) }, \label{zpm}
\end{equation}
\label{sphfreq}
\end{subequations}
where $L_z$ is the $z$-component of particle's angular momentum and $E$ is its energy per unit rest mass, which can be explicitly expressed as the functions of \{$r_s$, $a$, $Q$\} parameters (see Equation (16) in \cite{RMCQG2019}). The definitions of the Elliptic integrals are \cite{Grad}
\begin{subequations}
\begin{eqnarray}
F\left( \varphi, p^2 \right)=&&  \int_{0}^{\varphi}\frac{{\rm d}\alpha}{\sqrt{1-p^2 \sin^2 \alpha}}, \label{elliptica} \\
K\left( \varphi, p^2 \right)=&&  \int_{0}^{\varphi}\sqrt{1-p^2 \sin^2 \alpha}\cdot {\rm d}\alpha, \label{ellipticb} \\
\Pi \left(q^2, \varphi, p^2 \right)=&& \int_{0}^{\varphi}\frac{{\rm d}\alpha}{\left( 1- q^2 \sin^2 \alpha \right) \sqrt{1-p^2 \sin^2 \alpha}}. \label{ellipticc}
\end{eqnarray}
\end{subequations}

\begin{figure}[H]
\centering
\mbox{
 \subfigure[]{
\includegraphics[scale=0.35]{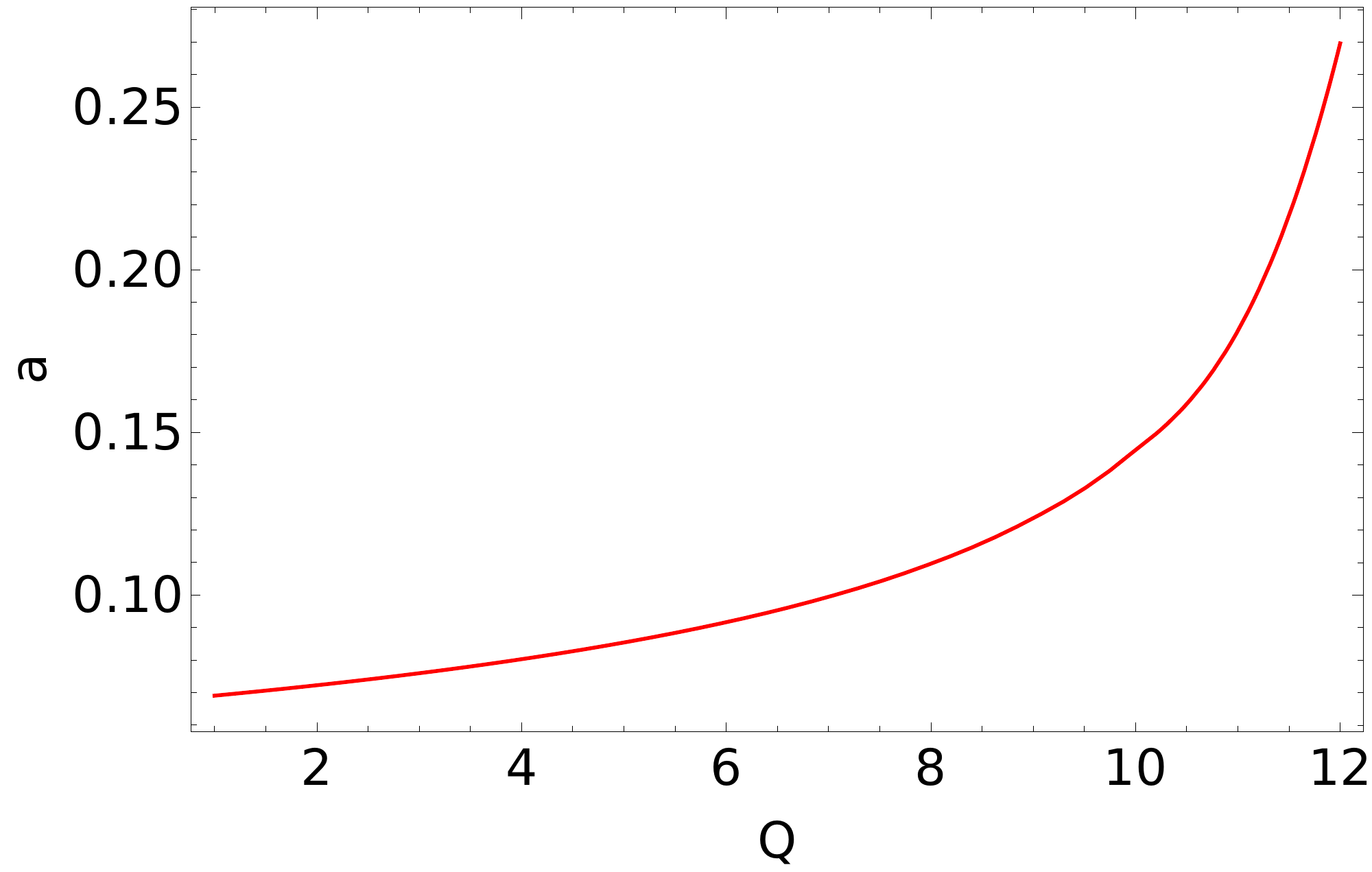}}
\hspace{0.7cm}
\subfigure[]{
\includegraphics[scale=0.35]{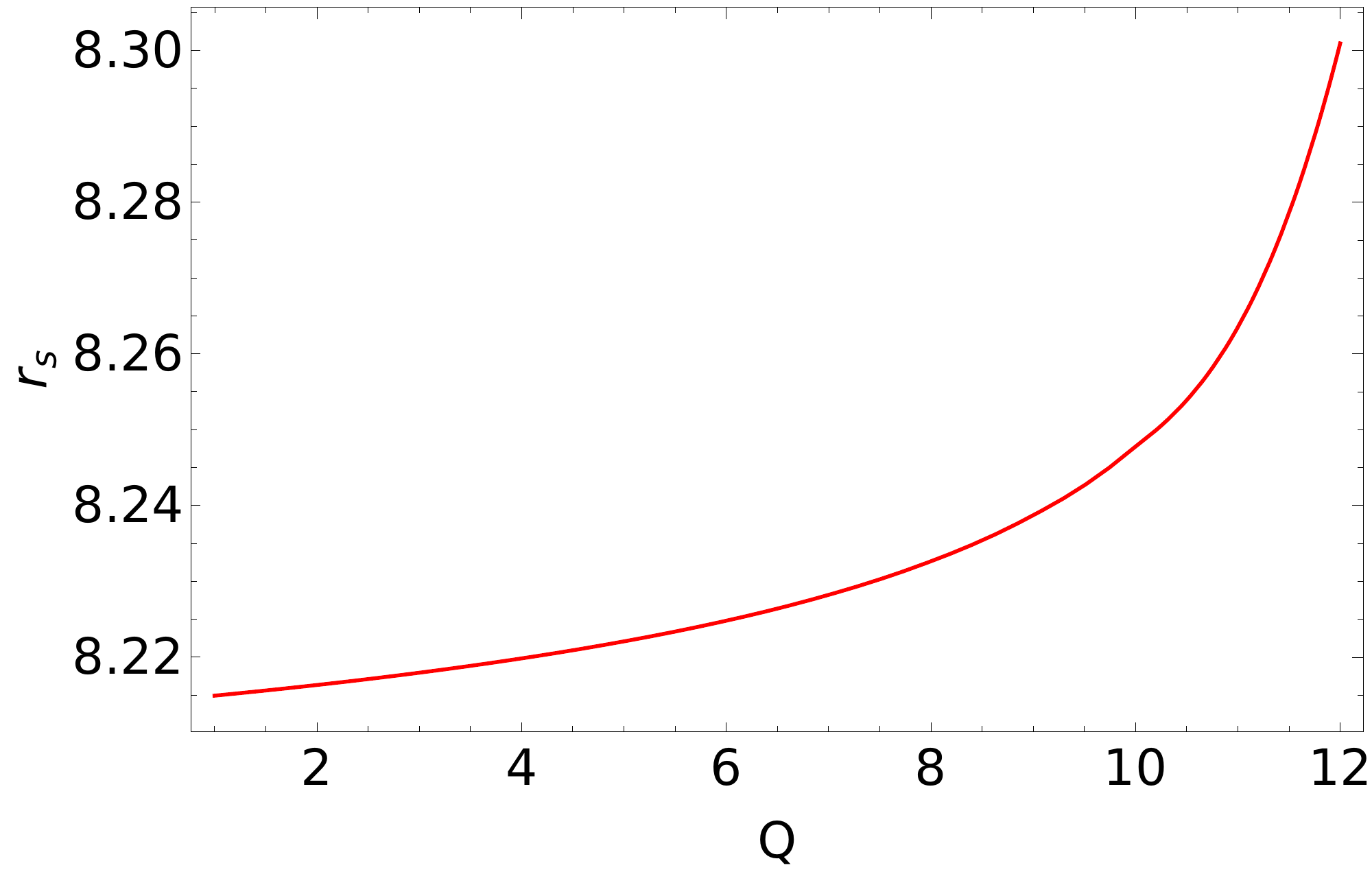}}}
\caption{\label{Qsolplot}The figure shows the solutions of spherical orbit parameters \{$r_{s0}$, $a_0$, $Q_0$\} for QPO frequencies of 1H 0707-495 in (\textbf{a}) ($Q$, $a$), and in (\textbf{b}) ($Q$, $r_s$) plane. }
\end{figure}

\begin{figure}[H]
\centering
\mbox{
 \subfigure[]{
\includegraphics[scale=0.32]{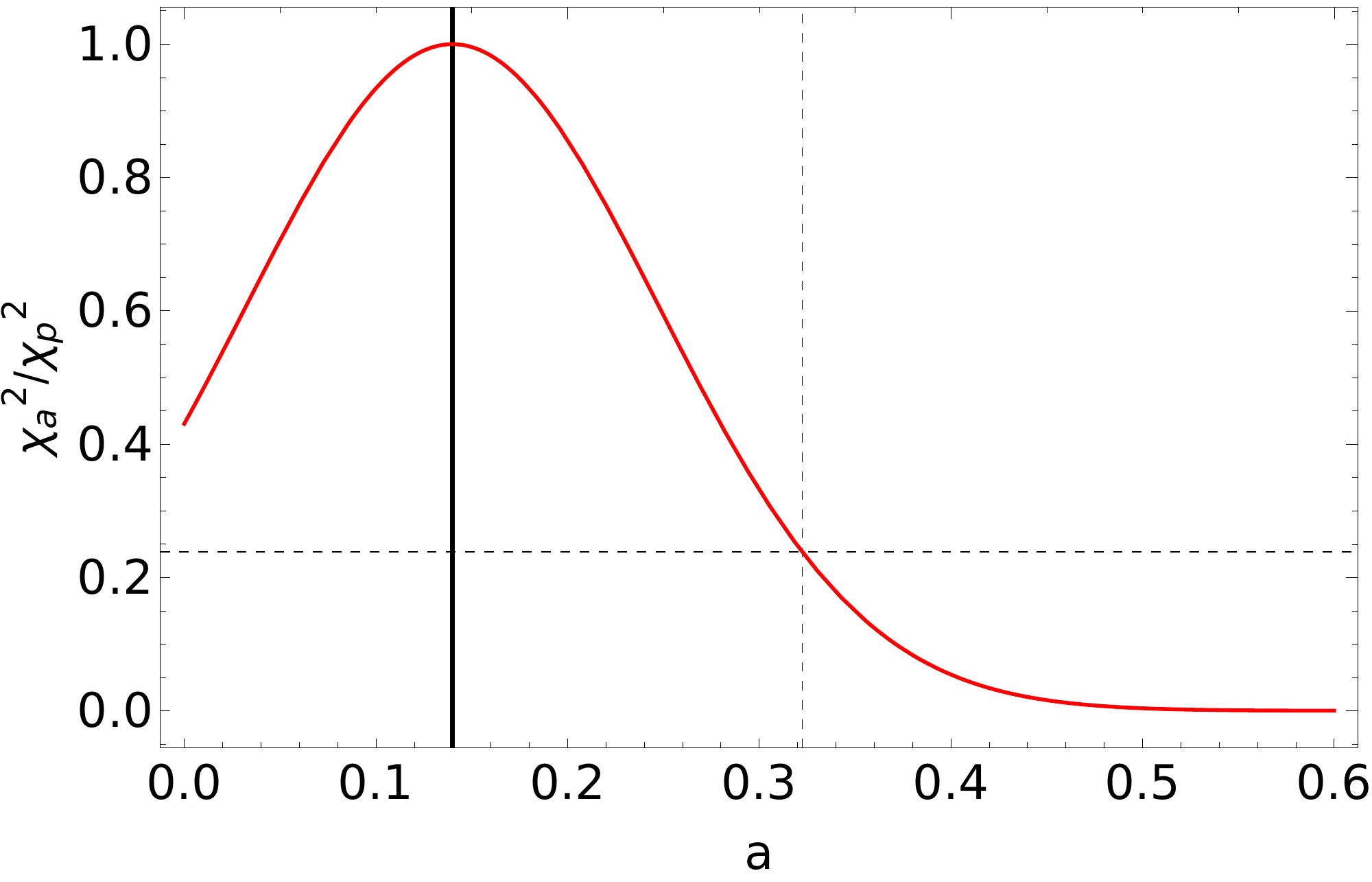}\label{chisqraplot}}
\hspace{1.2cm}
\subfigure[]{
\includegraphics[scale=0.32]{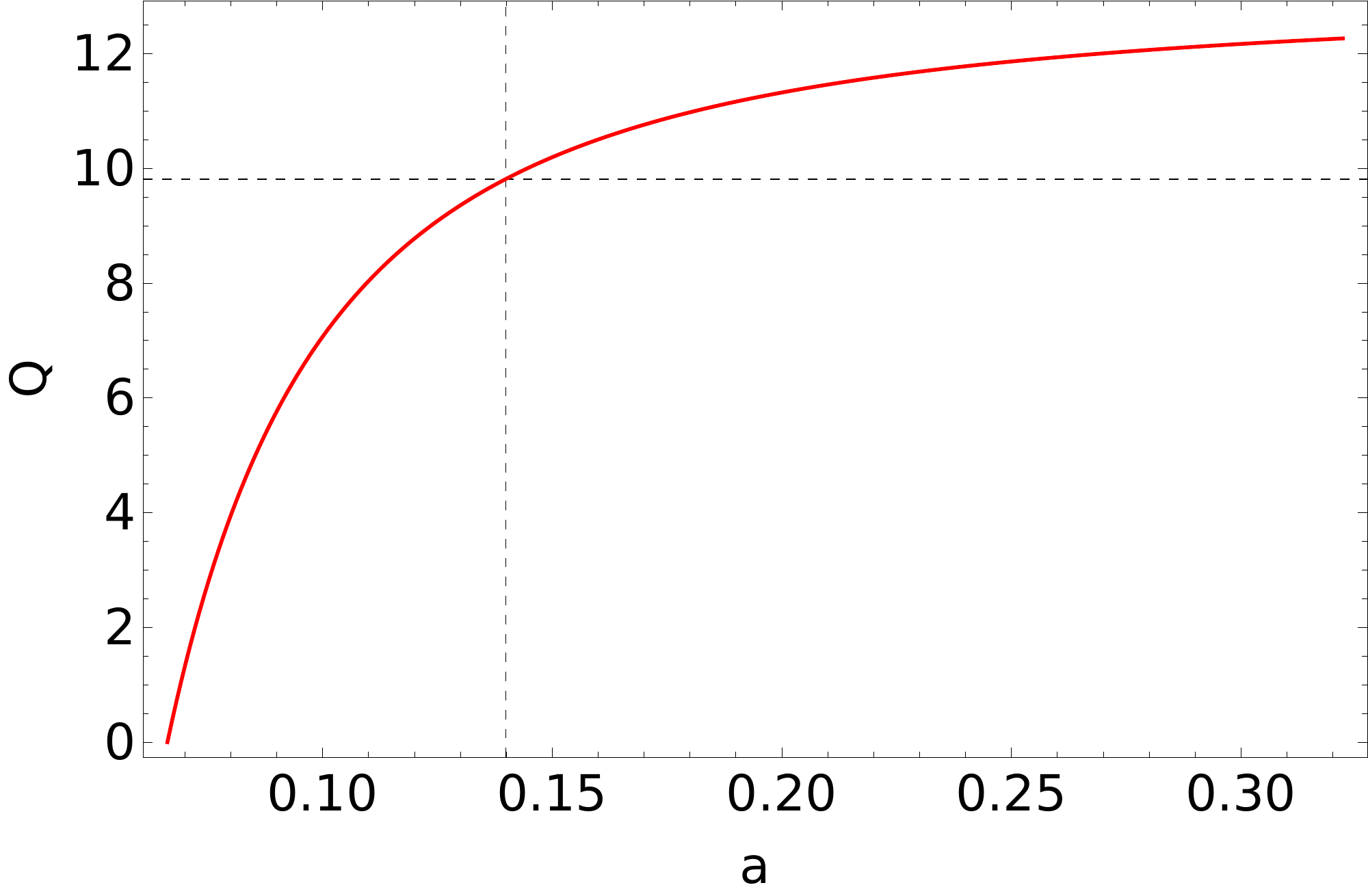}\label{aQplot}}}
\mbox{
\hspace{3.8cm}
 \subfigure[]{
\includegraphics[scale=0.32]{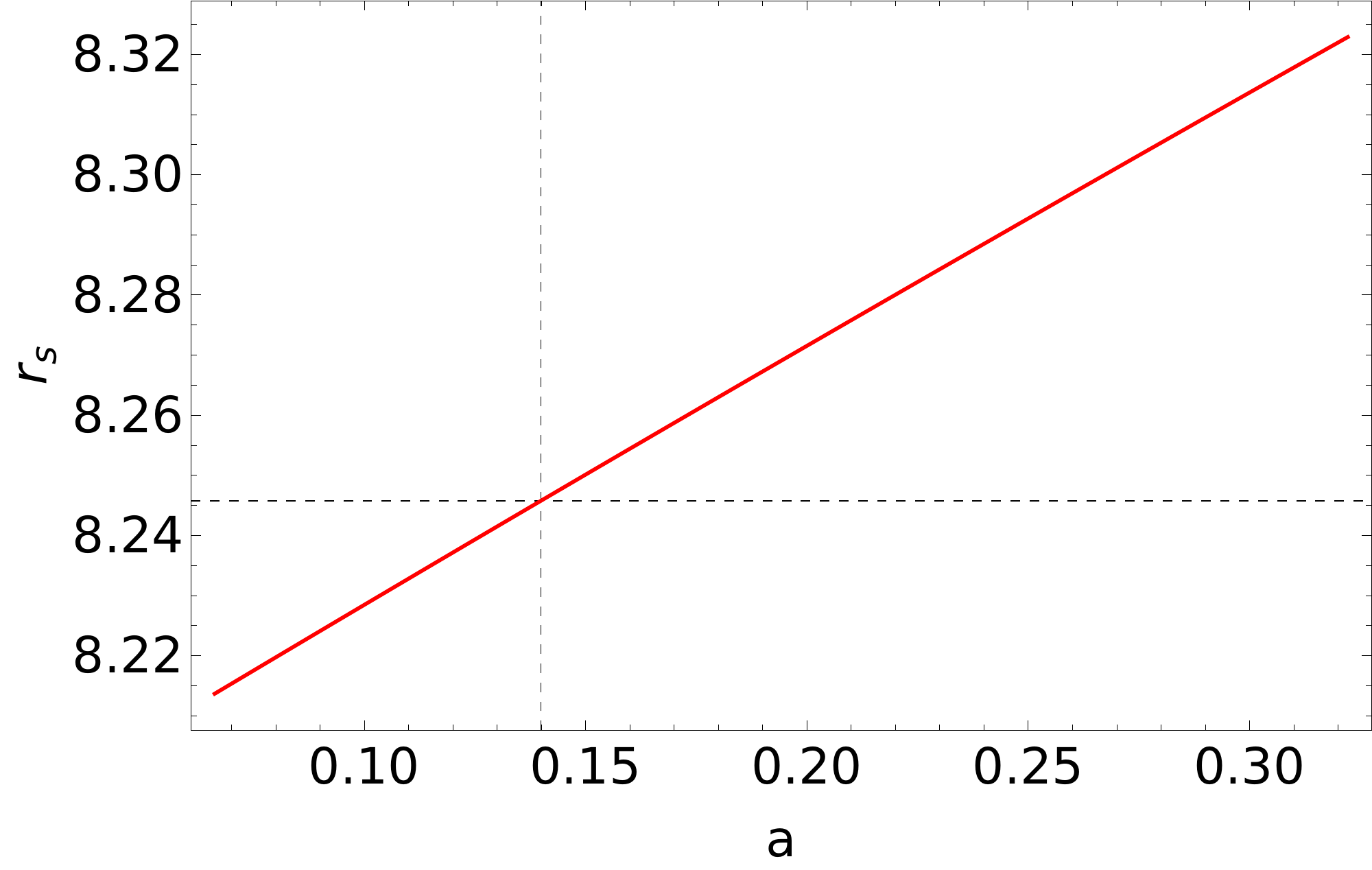}\label{arsplot}}}
\caption{\label{chisqrplot}The figure shows (\textbf{a}) $\chi^2_a/\chi^2_p$ function for $a$, where the vertical solid black curve depicts $a_p$ and the vertical dashed black curve encloses the 95\% probability region, (\textbf{b}) the range of $Q$ and (\textbf{c}) $r_s$ corresponding to the $2\sigma$ region of $a$, where the vertical dashed  black curves mark \{$a_p$ ,$r_{sp}$, $Q_p$\}. }
\end{figure}

We discuss our results below:
\begin{enumerate}
\item We explore the parameter space ($r_s$, $a$, $Q$) for the spherical orbits. Since there are two input QPO frequencies, we first vary the $Q$ value to find various solutions of \{$r_s$, $a$\} by solving equations \{$\nu_{\phi}=\nu_1$, $\nu_{pp}=\nu_2$\} as per GRPM. $Q=13$ is at the limit of astrophysically allowed bound orbits, Equation \eqref{boundcnd}; $Q<13$ in the case of 1H 0707-495. The $Q=13$ orbit is an unstable orbit very close to the separation of bound and unbound (called a separatrix orbit), and such an unstable orbit is not relevant to our study; hence, we fix our parameter exploration between Q= 1 and 12. In Figure \ref{Qsolplot}, we have shown these solutions in the ($Q$, $a$) and ($Q$, $r_s$) planes. 
\item Next, we fix $Q=\{1, 4, 8, 12\}$ and find the errors in the \{$r_s$, $a$\} parameters using the method described in Section \ref{errorestimate}. The range of \{$r_s$, $a$\}, resolution taken in the simulations, along with the exact solutions and their errors obtained by fitting $\mathcal{P}_{1} \left( r_s\right)$ and $\mathcal{P}_{1} \left( a\right)$ are summarized in Table \ref{1H07sphtable1}. 
\item The ranges of \{$a$, $r_s$, $Q$\}, shown in Table \ref{1H07sphtable1} and Figure \ref{Qsolplot}, span the complete parameter volume for QPO frequencies of 1H 0707-495. As the spin of the black hole does not change in the timescale of a few months or years, we need to find the most probable value of spin. We first find the variance of $\mathcal{P}_{1} \left( a\right)$ with respect to the exact solution of $a$ for each $Q$, given in Table \ref{1H07sphtable1}, which is given by 
\begin{subequations}
\begin{equation}
{\sigma_{ai}}^2=\int_{0}^{0.9} \left(a-a_{0i} \right)^2 \mathcal{P}_{1i} \left( a\right) da,
\end{equation}
where $\mathcal{P}_{1i} \left( a\right)$ is the probability density ditribution in $a$ parameter space for each value of $Q$. We have summarized the values of $\sigma_a$ for each $Q$ in Table \ref{1H07sphtable1}. We then minimize the likelihood function
\begin{equation}
L\left(a\right)= \sum_i^4 \frac{\left(a-a_{0i} \right)^2}{{\sigma_{ai}}^2},
\end{equation}
to obtain the most probable value of the spin given by
\begin{equation}
a_p=\frac{\sum_i^4 \left(a_{01}/ {\sigma_{ai}}^2\right)}{\sum_i^4 \left( 1/{\sigma_{ai}}^2\right)}.
\end{equation}
\end{subequations}
We find the peak value to be $a_p=0.139$ for 1H 0707-495, and corresponding solution of \{$r_s$, $Q$\} for the QPO frequencies is \{$r_{sp}=8.246$, $Q_p=9.814$\}.
\item Next, we obtain the $\chi^2_a$ distribution function of $a$ given by
\begin{equation}
\chi^2_a\left(a\right)=\exp\left[ -L \left(a\right)\right].
\end{equation}
A plot of $\chi^2_a/\chi^2_p$ is shown in Figure \ref{chisqrplot}a, where $\chi^2_p=\chi^2_a\left(a_p\right)$. We obtain the $2\sigma$ errors with respect to $a_p$ by normalizing the $\chi^2_a\left(a\right)$ function and obtain $0.139_{-0.139}^{0.183}$, where the region of 95\% probability is indicated by the vertical dashed line in Figure \ref{chisqrplot}a. We also show the range of $r_s$ and $Q$ in Figure \ref{chisqrplot}b,c within the $2\sigma$ region of $a$, as seen in Figure \ref{chisqrplot}a, where the parameter ranges are $r_s=(8.214-8.323)$ and $Q=(0.0001-12.264)$.
\item Hence, we conclude that the spherical orbits, close to the black hole in the region, $r_s=\left(8.214-8.323\right)$ with $Q$ values between $\left(0.0001-12.264\right)$, are possible sources of the QPO frequencies observed in 1H 0707-495, while the most probable spin value to be $a_p=0.139_{-0.139}^{0.183}$ with $2\sigma$ confidence.
\end{enumerate} 

\begin{table}[H]
\caption{The table summarizes results of spherical orbit parameter solution, \{$r_s$, $a$\}, and corresponding errors for X-ray QPOs in NLSy1 1H 0707-495. The columns provide the range of parameter volume taken for \{$r_s$, $a$\} by fixing $\{Q=1, 4, 8, 12\}$, the chosen resolution to calculate the normalized probability density at each point inside the parameter volume, the exact solutions, the results of the model fit to the integrated profiles, and variance $\sigma_a$. The mass of the black hole is fixed to $M_{\bullet}=5.2 \times 10^6 M_{\odot}$ \cite{Panetal2016}. }
\centering
 \tablesize{\footnotesize}
 \scalebox{0.95}[0.95]{
\begin{tabular}{cccccccccc}
\toprule
\textbf{$Q$} & \textbf{$r_s$ Range} & \textbf{Resolution} & \textbf{Exact Solution}& \textbf{Model Fit} & \textbf{$a$ Range} & \textbf{Resolution} & \textbf{Exact Solution} & \textbf{Model Fit} & \textbf{$\sigma_a$}\\
&  & \textbf{$\Delta r_s$} & \textbf{$r_{s0}$} &  &  & \textbf{$\Delta  a$} & \textbf{$a_0$} & &\\
\midrule
1 & 6.5--9.5 & 0.01 & 8.215 & $8.215_{-0.354}^{+0.118}$ & 0--0.9 & 0.001 & 0.069 &  $0.069_{-0.069}^{+0.28}$  & 0.290\\\midrule
4 & 6.5--9.5 & 0.01 & 8.219 & $8.219_{-0.331}^{+0.127}$ & 0--0.9 & 0.001 & 0.080 & $0.080_{-0.080}^{+0.316}$ &  0.317\\\midrule
8 & 6.5--10 & 0.01 &  8.233 & $8.233_{-0.278}^{+0.157}$ & 0--0.9 & 0.001 & 0.109 &$0.109_{-0.109}^{+0.366}$ & 0.348\\\midrule
12 & 6.5--10 & 0.01 &  8.301 & $8.301_{-0.196}^{+0.231}$ & 0--0.9 & 0.001 & 0.269 & $0.269_{-0.269}^{+0.127}$& 0.277\\
\bottomrule
\end{tabular}}
\label{1H07sphtable1}
\end{table}

\section{Relativistic Jet Model for the Optical and $\gamma$ Ray QPOs}
\label{Jetmodel}

In a simple kinematic approach inspired by the lighthouse model \cite{CK1992,MohanMangalam2015,Mangalam2018}, the basic periodicity is set by
\begin{eqnarray}
T_F=&&30.93 \left( r_F^{3/2}+a \right) (1+z) m_6 \ \mathrm{s}, \nonumber  \\
\simeq&& 35.8 \left(\frac{r_F}{100} \right)^{3/2} \left(1+z \right) m_8 \ \mathrm{days},  \label{timescaleformula}
\end{eqnarray}
where $m_6=M_{\bullet}/\left( 10^6 M_{\odot}\right)$ and $m_8=M_{\bullet}/\left( 10^8 M_{\odot}\right)$ and $r_F$ is the radius of the footpoint of the magnetic field anchored in the equatorial plane. An important radius is the light cylinder radius, which given in geometrical units is $r_L =r_F^{3/2} +a$. The plasma is expected to relativistically follow the field lines upto the light cylinder rigidly beyond which the field lines would be bend. A reasonable estimate of the cylindrical radius of the plasma motion is expected to be typically $r_0\equiv \chi r_L$ where $\chi=\left(0.1-10\right)$. Taking~an angular momentum conservation beyond the Alfven radius, $r_A =x_A r_L$, where $x_A <1$, will lead to $r_0^2 \Omega_0^2 = x_A^2 r_L^2 \Omega_F^2$, setting an observed periodicity of $T_0 = \left(\chi /x_A\right) T_F$. The value of $\left(\chi /x_A\right)$ depends on details of the relativistic MHD models and $x_A$ is determined by the relativistic Bernoulli equation, but a range of $\left(\chi /x_A\right) = 1-20$ is reasonable \cite{MohanMangalam2015}. This is illustrated by estimating $T_F(r_F)$ for the range of $r_F=(30-80)$ (see Table \ref{AGNjetQPO}); we see that the observed $\displaystyle{\frac{T_0}{T_F(r_F=50)}}$ is in the range of $\left(1-20\right)$.

This agreement motivates the study of the plasma motion in the background of relativistic MHD models, and its comparison with fits to the light curves in the future. Another clue of the jet physics will come from polarization models, as evidenced by the promising but simplistic cylindrical relativistic polarization signatures of the EVPA, DOP, and Doppler boosted flux profiles, as predicted by \cite{Mangalam2018}; this will be an additional and useful tool to extract jet properties by doing detailed fits to polarization observations. There is an oscillatory behavior seen in both $\gamma$-ray and optical light curves \cite{Sandrinelli2014,Sandrinelli2016a,Sandrinelli2016b,Ackermann2015,Sandrinelli2018,Sandrinelli2017,Bhatta2019,Guptaetal2019} that supports the above trend. There is also evidence of the radio structure that is supported by the basic model of \cite{MohanMangalam2015} as observed by \cite{Anetal2020,Mohanetal2016}.

\begin{table}[H]
\caption{A list of statistically significant QPOs detected in the $\gamma$ ray and optical energy bands in BL Lacertae type of AGN, along with their redshifts and black hole masses. The theoretical timescales are calculated, using Equation \eqref{timescaleformula}, such that the lower and upper limit correspond to $r_F=30$ and $r_F=80$ respectively. The lower-case letters (a to m) are links to references given in the last column.}
\centering
 \tablesize{\footnotesize}
 %% You can specify the fontsize here, e.g., \tablesize{\footnotesize}. If commented out \small will be used.
\begin{tabular}{cccccccccc}
\toprule
\textbf{\#} & \textbf{Source}  & \textbf{$z$}  &\textbf{Log$\left(M_{\bullet}/M_{\odot}\right)$}  & \textbf{Energy Band} & \textbf{QPO Period} & \textbf{$T_F$}& \textbf{References}\\
& & & & & \textbf{$T_0$ (Days)} & \textbf{ (Days)}& & \\
\midrule
1. & PKS 2155-304 & 0.116 $^{\rm a}$ & 8.7 $^{\rm b}$ & 100 MeV--300 GeV & 620 $\pm$ 41 $^{\rm c}$ & 33--143 &  \cite{Ackermannetal2015} $^{\rm a}$, \cite{Chen2018} $^{\rm b}$, \cite{Sandrinelli2014,Sandrinelli2016a,Sandrinelli2018} $^{\rm c}$ \\
&  & & & 100 MeV--300 GeV  & 612 $ \pm $ 42 $^{\rm d}$ & & \cite{Tarnopolski2020} $^{\rm d}$\\
 &  & & & R (optical)  & 315 $ \pm $ 25 $^{\rm c}$ & & \\
 \midrule
2. & PG 1553+113  & 0.36 $^{\rm e}$ & $\sim$ 8 $^{\rm f}$& 100 MeV--300 GeV & 780 $ \pm $ 63 $^{\rm g}$ & 8--35 &  \cite{Chen2018} $^{\rm e}$, \cite{Tavani2018} $^{\rm f}$, \cite{Ackermann2015,Sandrinelli2018} $^{\rm g}$ \\
 & & &  & R (optical) & 810 $ \pm $ 52 $^{\rm g}$ &  & \\\midrule
3. & PKS 0537-441 & 0.892 $^{\rm h}$ & 8.56 $^{\rm i}$ & 100 MeV--300 GeV & 280 $ \pm $ 39 $^{\rm j}$ & 40- 176  & \cite{Ackermannetal2015} $^{\rm h}$, \cite{Chen2018} $^{\rm i}$, \cite{Sandrinelli2016b} $^{\rm j}$\\
 & & &  & R (optical) & 148 $ \pm $ 17 $^{\rm j}$ & & \\\midrule
4.  & BL Lac & 0.0686 $^{\rm k}$ & 8.21 $^{\rm l}$ & 100 MeV--300 GeV & 680 $ \pm $ 35 $^{\rm m}$ & 10--44 & \cite{Ackermannetal2015} $^{\rm k}$, \cite{Chen2018} $^{\rm l}$, \cite{Sandrinelli2017,Sandrinelli2018} $^{\rm m}$ \\
& & &  & R (optical) & 670 $ \pm $ 40 $^{\rm m}$ & & \\
\bottomrule
\end{tabular}
\label{AGNjetQPO}
\end{table}

\section{Relativistic Orbit Model (ROM) and PSD Shape}
\label{PSDmodel}
The X-ray timing analysis of NLSy1 galaxies has been proven to be an essential tool for probing the emission region and the underlying mechanism of the variability process of the X-ray flux in these sources. The shape of the power spectral density is found to have a shape which is well fit by a bending power-law model given by \cite{McHardy2004}
\begin{equation}
\mathcal{P}_s\left(\nu \right)=P_0 \left(\frac{\nu}{\nu_{b}} \right)^{-\alpha_l}\left[1+ \left(\frac{\nu}{\nu_{b}} \right)^{\left(\alpha_h  -\alpha_l \right)} \right]^{-1}, \label{bendlaw}
\end{equation}    
where $P_0$ is the normalization constant, and $\alpha_l$, $\alpha_h$ are the PSD slopes
below and above the break frequency, $\nu_b$. The power density spectrum shows that the low-frequency power spectrum is significantly flatter ($\alpha_l \sim1$) than the high-frequency power spectrum ($\alpha_h > 2$). The break frequencies were found to be near $\nu_b \sim6.7 \times 10^{-6}$Hz for PKS 0558-504 \cite{Papadakis2010} and $\nu_b \sim8 \times 10^{-4}$Hz for NGC 4051 \cite{McHardy2004}.

Here, we present a plausible relativistic orbit model to generate such a power density spectrum. As argued before, the non-equatorial orbits, such as spherical orbits, are the natural consequence of the axisymmetry of the Kerr space-time \cite{Carter1968,RMCQG2019}. We assume that inside a spherical corona region of relativistic electrons (the inner corona, IC, $r_M<r<r_I$), existing inside the radius of innermost stable circular orbit (ISCO) (see Figure \ref{modelplot}), the particles are in non-equatorial orbits. The thin accretion disk spans the region outside ISCO, where the fluid motion is confined to the equatorial plane. We also assume that an outer corona region (OC, $r_I<r<r_X$) of relativistic particles envelopes this accretion disk, lying almost in the equatorial plane (see Figure \ref{modelplot}). The energy per unit rest mass of these relativistic particles, $E$, orbiting in the equatorial circular trajectories, is given by \cite{Bardeen1972}
\begin{equation}
E\left(r, a \right)=\frac{r^{2}-2r + a\sqrt{r}}{r\left( r^{2}- 3r + 2a\sqrt{r}\right) ^{1/2}}. \label{Ecirc}
\end{equation}

\begin{figure}[H]
\centering
\mbox{
\subfigure[]{
\includegraphics[scale=0.32]{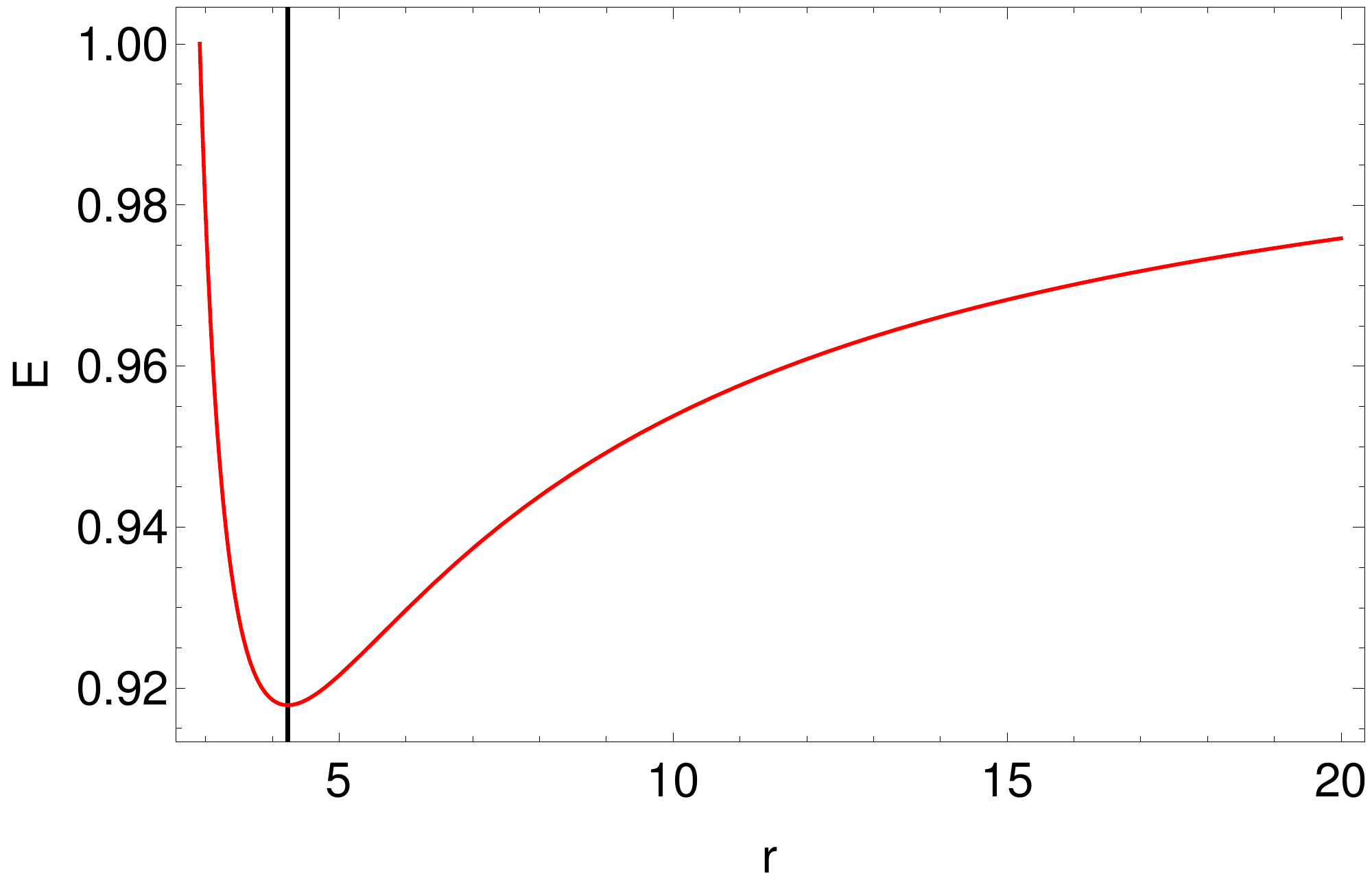}\label{plotEra05Q0}}
\hspace{0.4cm}
\subfigure[]{
\includegraphics[scale=0.32]{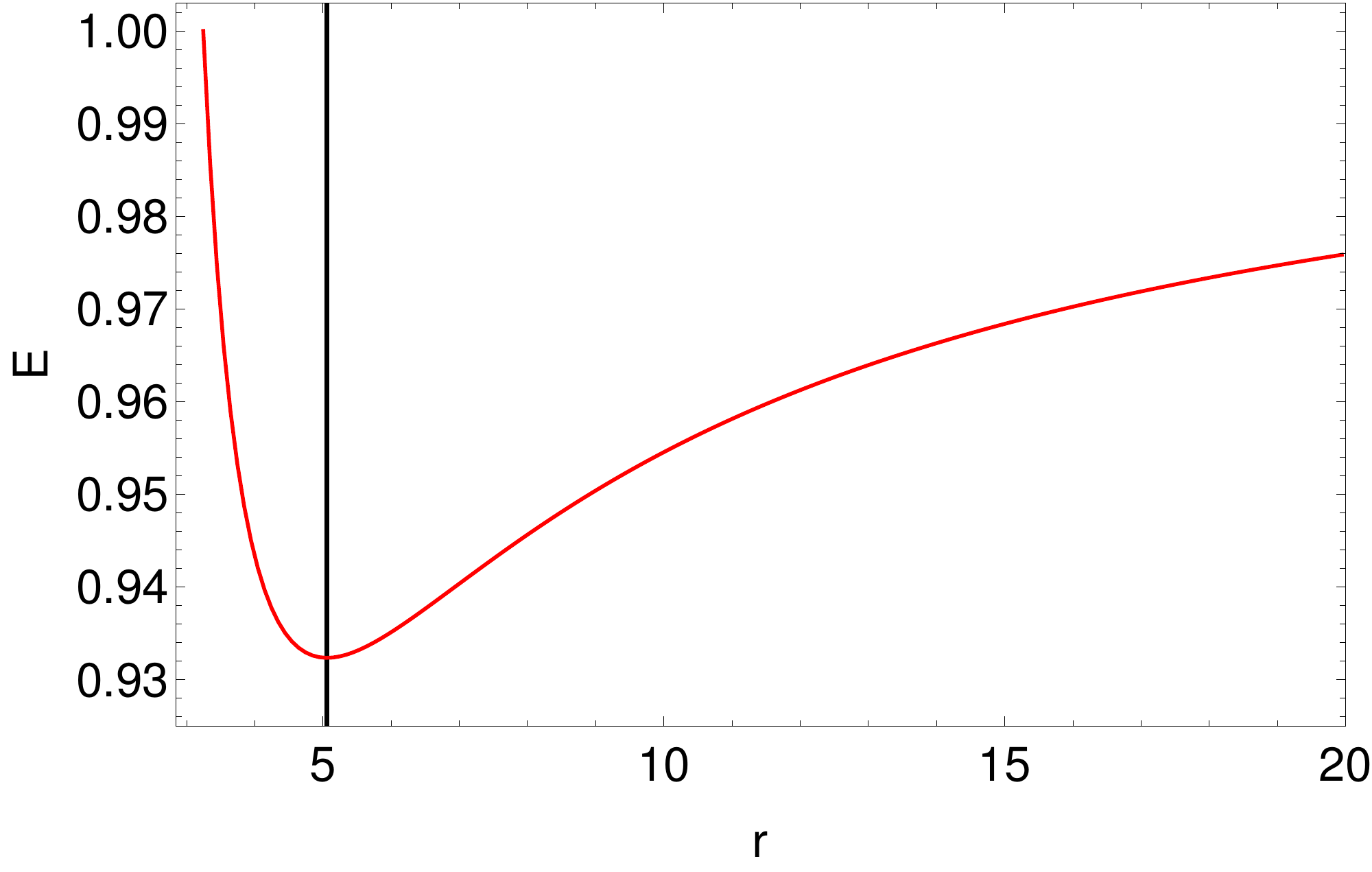}\label{plotEra05Q8}}}
\caption{\label{Erplot}The figure shows $E\left(r \right)$ as a function $r$ for (\textbf{a}) the equatorial circular orbits ($Q=0$) and for (\textbf{b}) the spherical orbits with $Q=8$, where $a=0.5$. The vertical black curves correspond to the innermost stable circular orbit (ISCO) and to the innermost stable spherical orbit (ISSO) for $Q=8$.} 
\end{figure}

We see that $E$ increases with $r$ outside ISCO, and it decreases with $r$ inside ISCO, where it has minima at the ISCO radius; see Figure \ref{Erplot}a. The stable circular orbits exist outside the ISCO radius, whereas the unstable circular orbits are found inside the ISCO radius. The mechanical energy per unit rest mass of the relativistic plasma, $E$, orbiting in the spherical trajectories, is given by \cite{RMCQG2019}
\begin{equation}
 E\left(r_s, a, Q \right)= \frac{\left\lbrace \splitfrac{2a^4 Q + \left( r_s-3\right) \left( r_s-2 \right)^2 r_{s}^4 -a^2 r_s \left[r_{s}^{2}\left(3r_s -5 \right) + Q \left(r_s\left(r_s -4 \right) +5  \right)   \right]}{ -2 a \left[r_s \left(r_s -2 \right) +a^2  \right] \sqrt{a^2 Q^2 -r_s^3 Q \left(r_s -3 \right) + r_s^5} } \right\rbrace^{1/2}  }{r_{s}^2 \left[r_{s} \left( r_s-3\right)^2 -4a^2 \right]^{1/2} }, 
 \label{Ensph} 
\end{equation}
where $E$ increases with $r_s$ outside the innermost stable spherical orbit (ISSO), and it decreases with $r_s$ inside ISSO, where it has minima at the ISSO radius; see Figure \ref{Erplot}b. The stable spherical orbits, for a fixed $Q$, exist outside the ISSO radius, whereas the unstable spherical orbits are found inside the ISSO radius. 

\begin{figure}[H]
\begin{center}
\includegraphics[scale=0.32]{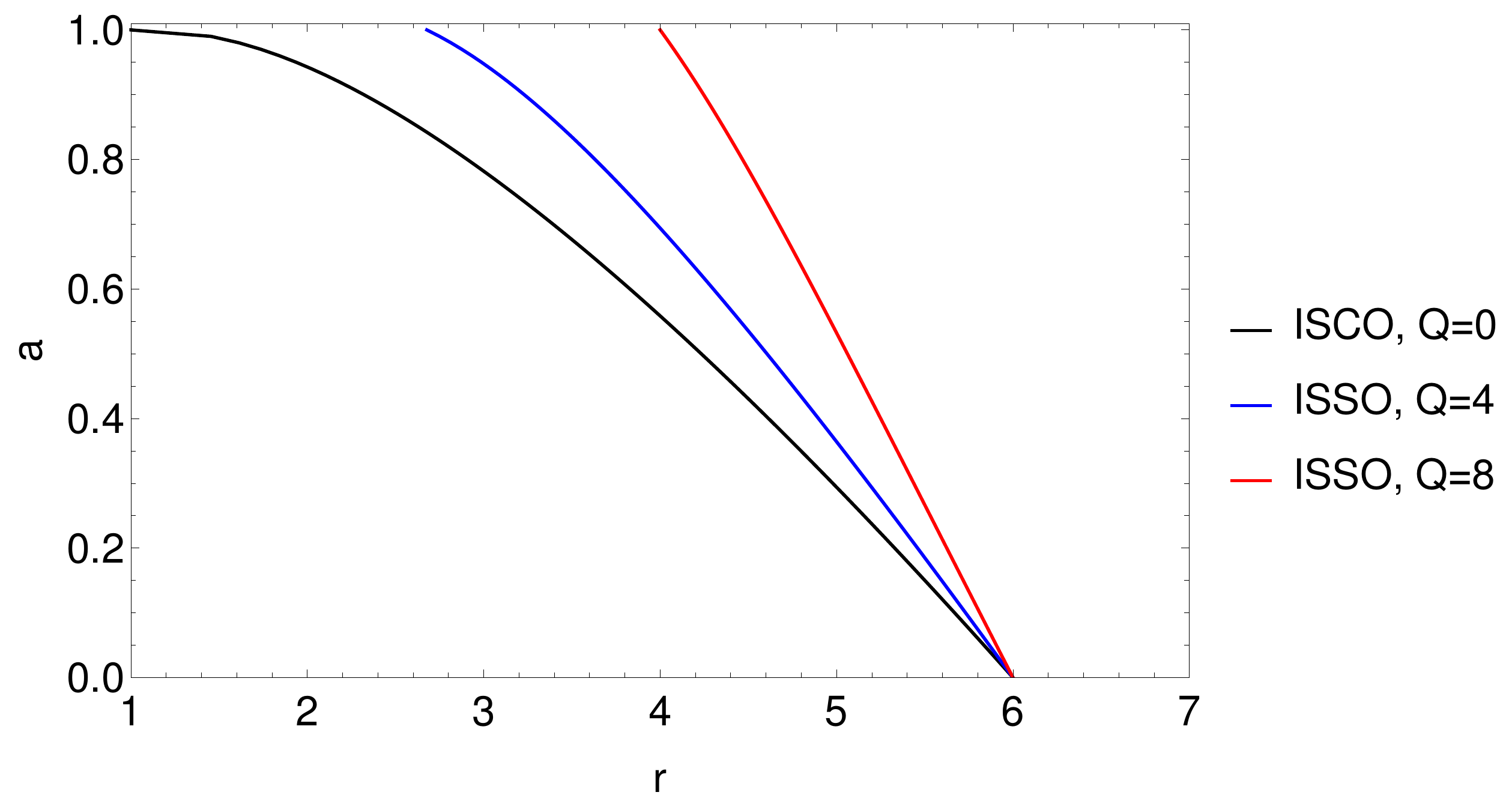}
\end{center}
\caption{\label{ISCOISSOraplot}The figure shows a comparison of the ISCO and ISSO radii in the ($r$, $a$) plane. The ISSO %Define, if appropriate.
	radius moves outwards as $Q$ increases.}
\end{figure}

A comparison of the ISCO and ISSO radii is shown in the ($r$, $a$) plane in Figure \ref{ISCOISSOraplot}, where we see that these radii converge to $r=6$ for the Schwarzschild black hole ($a=0$), as the spherical orbits or ISSOs are possible only outside a Kerr black hole ($a\neq0$) because of the axisymmetry of the Kerr space-time. Additionally, as the value of $Q$ increases, the ISSO radii move further away from the black hole as compared to the ISCO radius. This implies that one will always find the unstable circular and the unstable spherical orbits inside the ISCO radius.
\subsection{The ROM}
\begin{figure}[H]
\centering
\mbox{
\subfigure[]{
\includegraphics[scale=0.35]{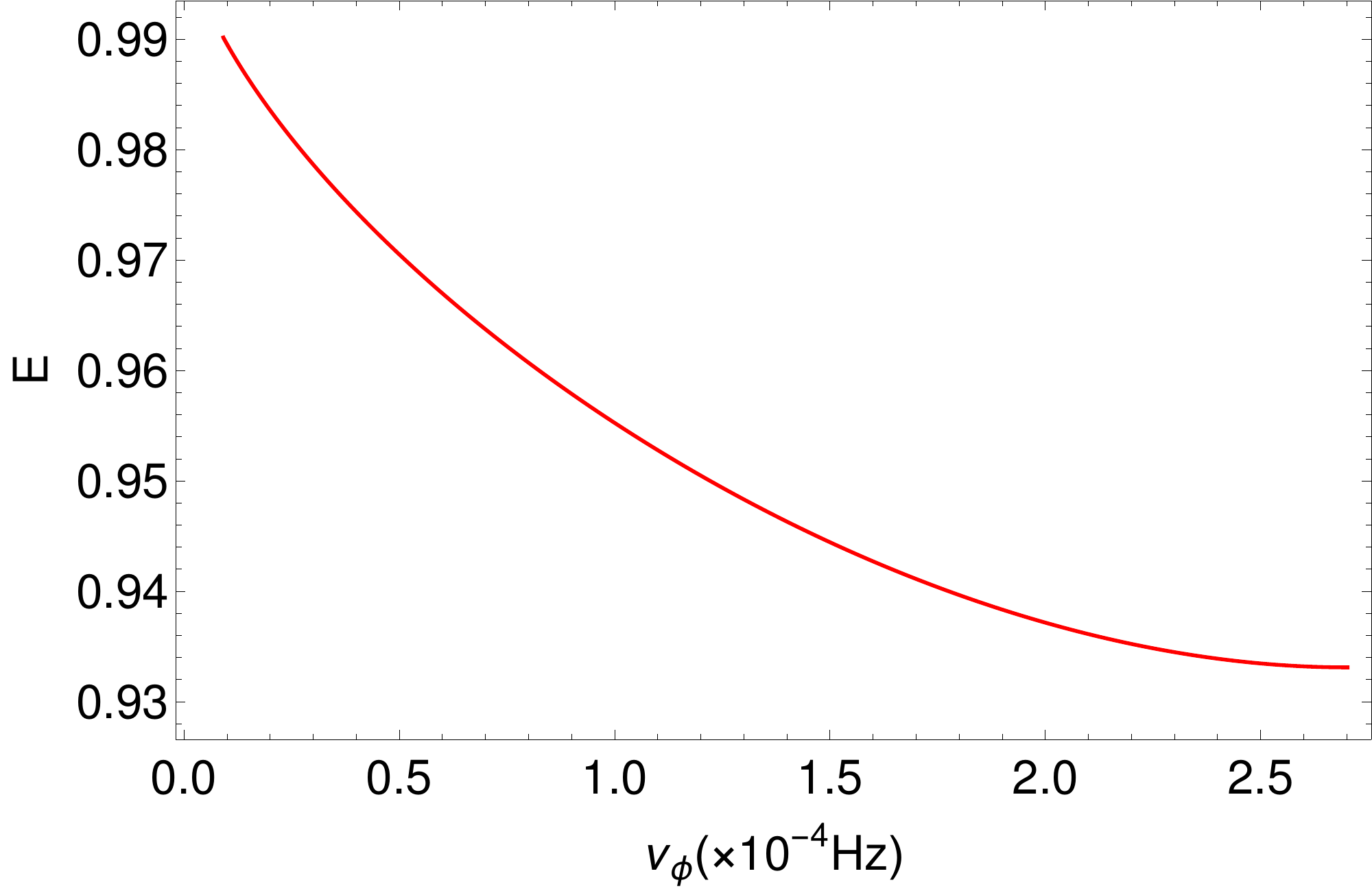}\label{freqEplotouta25Q0}}
\hspace{0.4cm}
\subfigure[]{
\includegraphics[scale=0.35]{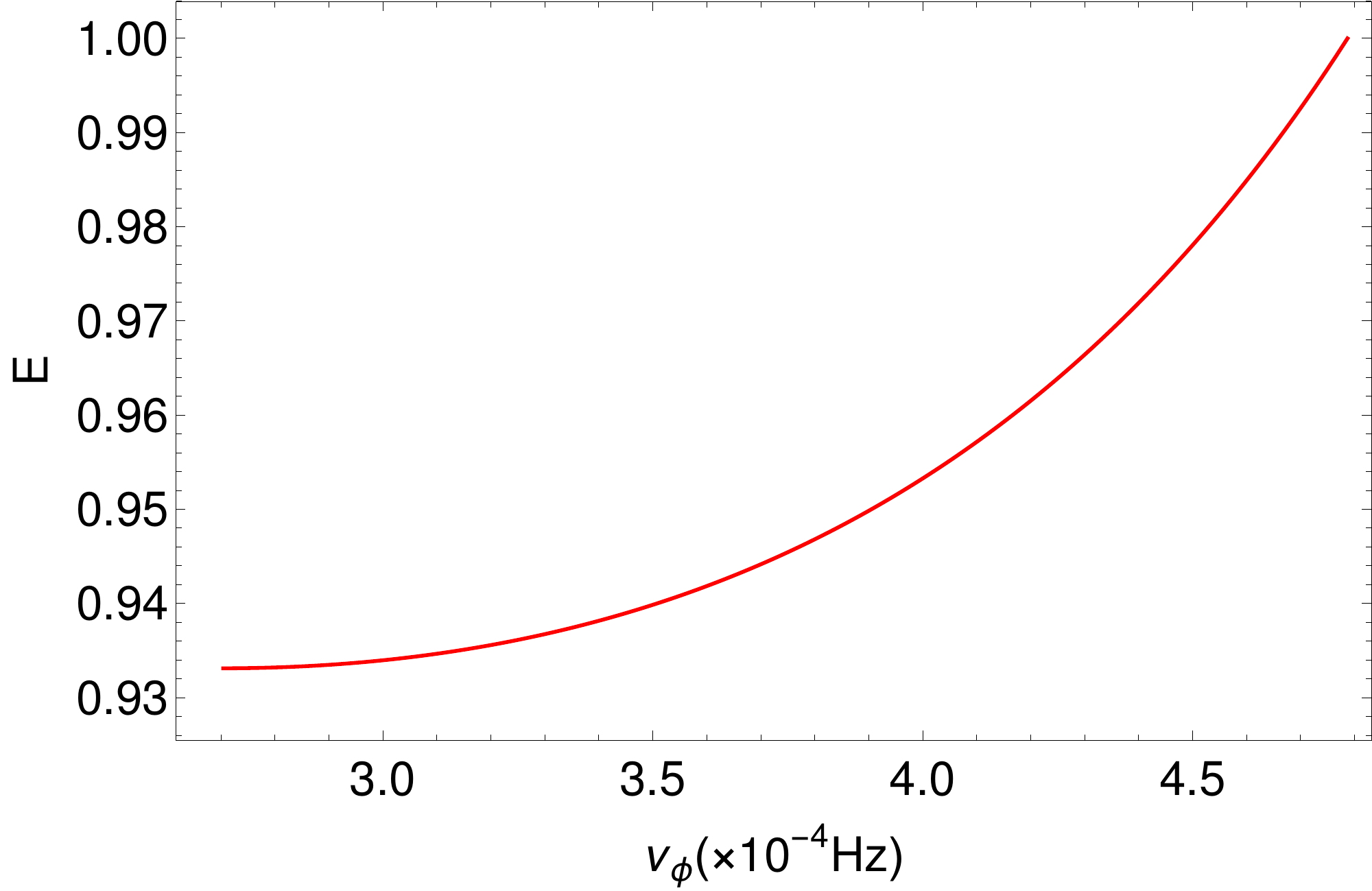}\label{freqEplotina25Q0}}}
\caption{\label{freqEplot}The figure shows $E$ as a function of $\nu_{\phi}$ for the circular orbits (\textbf{a}) outside $r_I$, and (\textbf{b}) inside $r_I$ for $a=0.25$ and $M_{\bullet}=10^{7}M_{\odot}$. The minima of $E$ is at $r_I$ in both diagrams.}
\end{figure}

The underlying assumptions of our model are:
\begin{enumerate}
\item We associate the temporal frequency, $\nu$, in the observed power spectral density with the fundamental azimuthal frequency of the particles orbiting in the circular orbits in the accretion disk outside ISCO, $r_I$, and both circular and spherical trajectories between $r_I$ and marginally bound spherical orbit (MBSO) radius, $r_M$. These frequencies are functions of the orbital radius, $r$ or $r_s$, (Equations \eqref{nuphicirc} and \eqref{nuphisph2}), and hence they are also fundamentally related to the mechanical energy of the orbit through Equations (\ref{Ecirc}) and (\ref{Ensph}).
\item We assume a prior distribution of the energy of particles (or electrons) given by a power-law
\begin{subequations}
\begin{eqnarray}
N\left(E\right)=&& A \left(\frac{E}{E_{I}}\right)^{-\alpha_1} , \ \ \ \mathrm{IC:\ radial \ range\ }  r_M < r< r_I, \\
               = && A \left(\frac{E}{E_{I}}\right)^{-\alpha_2}, \ \ \ \mathrm{OC:\ radial \ range \ } r_I < r< r_X.
\end{eqnarray}
\label{NE}
\end{subequations}
where $N\left(E\right)$ represents the number of particles having energy $E$, $\alpha_1$ and $\alpha_2$ are the power-law indices inside and outside $r_I$ respectively, $E_{I}$ is the particle energy at $r_I$, and  $A$ is the normalization constant. The energy distribution, $N\left(E\right)$ (Equation \eqref{NE}), is constructed so that it is continous at $r_I$. Assuming that the total number of particles are $N_0$ (however, the PSD solution is independent of this), we have the normalization condition given by
\begin{subequations}
\begin{equation}
\int N\left( E \right) d E = N_0,
\end{equation}
\begin{equation}
A \int_{E_{I}}^{1} \left(\frac{E}{E_{I}}\right)^{-\alpha_1} dE + A \int_{E_{I}}^{E_{X}} \left(\frac{E}{E_{I}}\right)^{-\alpha_2} dE =N_0,
\end{equation}
where the first and the second terms contribute for the regions inside and outside $r_I$ respectively, and $E_{X}$ corresponds to the energy of the particles at the outer radius of the equatorial circular accretion disk, $r_X$. Subsequently, we obtain
\begin{equation}
A\left(a, \alpha_1, \alpha_2 \right)=N_{0}\left[ \frac{\left(E_{I}\left(a \right)^{\alpha_1}-E_{I}\left(a \right) \right)}{\left(1- \alpha_1 \right)} + \frac{\left( E_{X}\left(a \right)^{\left(1-\alpha_2\right)}E_{I}\left(a \right)^{\alpha_2} - E_{I}\left(a \right)\right)}{\left(1- \alpha_2 \right)}\right]^{-1}. \label{A}
\end{equation}
We redefine $A\left(a \right)$ such that
\begin{equation}
A\left(a , \alpha_1, \alpha_2 \right)=N_0 B\left(a , \alpha_1, \alpha_2\right),
\end{equation}
where
\begin{equation}
B\left(a , \alpha_1, \alpha_2 \right)=\left[ \frac{\left(E_{I}\left(a \right)^{\alpha_1}-E_{I}\left(a \right) \right)}{\left(1- \alpha_1 \right)} + \frac{\left( E_{X}\left(a \right)^{\left(1-\alpha_2\right)}E_{I}\left(a \right)^{\alpha_2} - E_{I}\left(a \right)\right)}{\left(1- \alpha_2 \right)}\right]^{-1}. \label{B}
\end{equation}
\end{subequations}
\item We assume that the break frequency of the PSD corresponds to the temporal frequency at the ISCO~radius.
\item We also assume that the particle distrbution in the temporal frequency space, $F\left( \bar{\nu}\right)$, directly translates to the observed intensity for a given temporal frequency, so that the power density is given by $P\left(\bar{\nu}\right) \propto F \left( \bar{\nu}\right)^2 $.
\end{enumerate}
Next, we derive the distribution of the temporal frequency, $F\left(\bar{\nu}\right)$, as follows:
\begin{subequations}
\begin{equation}
\frac{dN\left(E\right)}{dE} = \frac{dF\left(\bar{\nu}\right)}{d\bar{\nu}} \frac{d\bar{\nu}}{dE},
\end{equation}
\begin{equation}
\Rightarrow \frac{dF\left(\bar{\nu}\right)}{d\bar{\nu}} =\frac{dN\left(E\right)}{dE}\frac{dE\left(\bar{\nu}\right)}{d\bar{\nu}}.
\end{equation}
\end{subequations}
We obtain $\left(dN\left(E\right)/dE\right)$ from Equation \eqref{NE}, and numerically obtain $E\left(\bar{\nu}\right)$ to derive $\left(dE\left(\bar{\nu}\right)/d\bar{\nu}\right)$, using Equations (\ref{nuphicirc}) and (\ref{Ecirc}) for circular and Equations (\ref{nuphisph2}) and (\ref{Ensph}) for spherical orbits. In Figure \ref{freqEplot}, we~have shown $E$($\bar{\nu}_{\phi}$), where it is clear that the behaviour of $E$($\bar{\nu}_{\phi}$) changes inside and outside $r_I$.  Outside $r_I$, Figure~\ref{freqEplot}a, the radius of the circular orbits goes from $r_I$ to $r_X \approx 50$ ($1$ keV$\approx \frac{G M_{\bullet}m_e}{r_X k_B}$), whereas inside $r_I$, Figure \ref{freqEplot}b, the radius of the circular orbits varies between $r_I$ and marginally bound circular orbit (MBCO), $r_M$, which corresponds to $E=1$.
Next, we obtain the temporal frequency distribution given by
\begin{subequations}
\begin{myequation}
F_1\left(\alpha_1, \alpha_2, \bar{\nu}, a, Q\right)= -\alpha_1 B\left(a , \alpha_1, \alpha_2\right) E_{I}\left(a\right)^{\alpha_1} \int^{\bar{\nu}}_{\bar{\nu}_{I}\left(a\right)} \Phi_1 \left(\alpha_1, \bar{\nu}^{'},a, Q \right)  d\bar{\nu}^{'} + N_I\left(\alpha_1, \alpha_2, a, Q \right), \ \ \mathrm{inside \ ISCO},  
\end{myequation}
\begin{myequation}
F_2\left(\alpha_1, \alpha_2, \bar{\nu}, a, Q\right) = -\alpha_2 B\left(a , \alpha_1, \alpha_2\right) E_{I}\left(a\right)^{\alpha_2} \int^{\bar{\nu}}_{\bar{\nu}_{X}\left(a\right)} \Phi_2\left(\alpha_2, \bar{\nu}^{'},a\right) d\bar{\nu}^{'} + N_X\left(\alpha_1, \alpha_2, a, Q \right), \ \ \mathrm{outside \ ISCO}, 
\end{myequation}
\label{Nnu}
\end{subequations}
where $\bar{\nu}_{I}\left(a\right)$ is frequency at $r_I$ and $\bar{\nu}_{X}\left(a\right)$ is frequency at $r_X$ (where the energy is $E_{X}$ defined in Equation \eqref{A}). $N_{I}\left(\alpha_1, \alpha_2, a, Q \right)$ and $N_X\left(\alpha_1, \alpha_2, a, Q \right)$ correspond to the number of particles having frequency at $r_I$ and $r_X$ respectively, where we have scaled the functions $F_1$, $F_2$, $N_I$, and $N_X$ by $N_0$. The expressions for the functions $\Phi_1 \left( \alpha_1, \bar{\nu}^{'},a, Q \right)$ and $\Phi_2 \left( \alpha_2, \bar{\nu}^{'}, a \right)$ are given by
\begin{subequations}
\begin{equation}
\Phi_1 \left(\alpha_1, \bar{\nu}^{'},a, Q \right) = \frac{1}{E\left(\alpha_1, \bar{\nu}^{'},a, Q \right)^{1+\alpha_1}} \frac{dE\left(\alpha_1, \bar{\nu}^{'},a, Q \right)}{d\bar{\nu}^{'}},
\end{equation}
\begin{equation}
\Phi_2 \left(\alpha_2, \bar{\nu}^{'},a \right) = \frac{1}{E\left( \alpha_2, \bar{\nu}^{'},a \right)^{1+\alpha_2}} \frac{dE\left(\alpha_2, \bar{\nu}^{'},a\right)}{d\bar{\nu}^{'}}.
\end{equation}
\end{subequations}

We scale the distribution functions, Equation \eqref{Nnu}, by $N_{I}$ for simplicity, which yields
\begin{subequations}
\begin{eqnarray}
f_1\left(\alpha_1, \alpha_2, \bar{\nu}, a, Q\right)=&& -\frac{V\left(a, \alpha_1, \alpha_2 \right)}{N_I\left(\alpha_1, \alpha_2, a, Q \right)}C_{1Ik}\left(\alpha_1, \bar{\nu} ,a, Q \right) + 1, \ \ \ \ \ \ \ \ \ \ \ \ \ \ \ \ \ \ \ \ \ \  \ \ \ \ \ \ \ \  \mathrm{inside \ ISCO}, \nonumber \\ \label{Nnu2} \\ 
f_2\left(\alpha_1, \alpha_2, \bar{\nu}, a, Q\right) =&& -\frac{W\left(a, \alpha_1, \alpha_2 \right)}{N_I\left(\alpha_1, \alpha_2, a, Q \right)}C_{2Xk}\left(\alpha_2, \bar{\nu} ,a\right) + n_X\left(\alpha_1, \alpha_2, a, Q \right), \ \ \ \ \mathrm{outside \ ISCO}, \nonumber \\ \label{Nnu3}
\end{eqnarray}
where 
\begin{equation}
C_{ijk}=\int_{\bar{\nu}_j}^{\bar{\nu}_k} \Phi_{i} \left( \bar{\nu}^{'}\right) d \bar{\nu}^{'},
\end{equation}
\begin{equation}
V\left(a, \alpha_1, \alpha_2 \right)=\alpha_1B\left(a , \alpha_1, \alpha_2\right)  E_I^{\alpha_1}, \ \ \ \ \ \ \ \ \ W\left(a, \alpha_1, \alpha_2 \right)=\alpha_2 B\left(a , \alpha_1, \alpha_2\right)  E_I^{\alpha_2}, 
\end{equation}
and
\begin{equation}
f_i \left(\alpha_1, \alpha_2,\bar{\nu}, a, Q\right)=\frac{F_i \left(\alpha_1, \alpha_2,\bar{\nu}, a, Q\right)}{N_I\left(\alpha_1, \alpha_2,a, Q \right)},  \ \ \ \ \ n_X\left(\alpha_1, \alpha_2,a,  Q \right)=\frac{N_X\left(\alpha_1, \alpha_2,a, Q \right)}{N_I\left(\alpha_1, \alpha_2, a, Q \right)} . \label{fnL}
\end{equation}
\label{Nnu1}
\end{subequations}

We employ the condition that $f_1\left(\alpha_1, \alpha_2,\bar{\nu}, a, Q\right)=f_2\left(\alpha_1, \alpha_2, \bar{\nu}, a, Q\right)$ at the frequency corresponding to \{$r_I$, $\bar{\nu}_{I}\left(a \right)$\}, which gives
\begin{subequations}
\begin{equation}
1=-\frac{W\left(a, \alpha_1, \alpha_2 \right)}{N_I\left(\alpha_1, \alpha_2, a, Q \right)} C_{2XI}\left(\alpha_2,a\right) + n_X\left(\alpha_1, \alpha_2, a, Q \right),
\end{equation}
or
\begin{equation}
N_I\left(\alpha_1, \alpha_2, a, Q \right)=-W\left(a, \alpha_1, \alpha_2 \right) C_{2XI}\left(\alpha_2, a\right) + N_X\left(\alpha_1, \alpha_2, a, Q \right). \label{EqNINL1}
\end{equation}
\end{subequations}

Next, we apply the normalization condition to the temporal frequency distribution given by
\begin{subequations}
\begin{equation}
F_1\left(\alpha_1, \alpha_2, \bar{\nu}_{M}, a, Q\right) + F_2\left(\alpha_1, \alpha_2, \bar{\nu}_{I}, a, Q\right)= 1,
\end{equation}
\begin{equation}
-V\left(a, \alpha_1, \alpha_2 \right) C_{1IM}\left(\alpha_1 ,a, Q \right) - W\left(a, \alpha_1, \alpha_2 \right) C_{2XI}\left(\alpha_2, a\right) + N_{I}\left(\alpha_1, \alpha_2, a, Q \right) + N_{X}\left(\alpha_1, \alpha_2, a, Q \right)=1, \label{normcondition}
\end{equation}
where $\bar{\nu}_{M}\left(a, Q\right)$ is the frequency at $r_M$. We solve Equations \eqref{normcondition} and \eqref{EqNINL1} together to obtain $N_{I}\left(\alpha_1, \alpha_2, a, Q \right)$ and $N_{X}\left(\alpha_1, \alpha_2, a, Q \right)$. Hence, the substitution of $N_{I}\left(\alpha_1, \alpha_2, a, Q \right)$ from Equation \eqref{EqNINL1} into Equation \eqref{normcondition} yields
\begin{eqnarray}
N_X\left(\alpha_1, \alpha_2, a, Q \right)=\frac{1}{2}\left[1+ V\left(a, \alpha_1, \alpha_2 \right) C_{1IM}\left(\alpha_1, a, Q\right)+2 W\left(a, \alpha_1, \alpha_2 \right) C_{2XM}\left(\alpha_2, a\right)\right]. \label{normalizeF}
\end{eqnarray}
\end{subequations}

By substituting Equation \eqref{normalizeF} back in Equation \eqref{EqNINL1}, we find
\begin{equation}
N_I\left(\alpha_1, \alpha_2, a, Q \right)= \frac{1}{2}\left[1+ V\left(a, \alpha_1, \alpha_2 \right) C_{1IM}\left(\alpha_1, a, Q\right)\right]. \label{EqNINL}
\end{equation}

Hence, we obtain $N_X\left(\alpha_1, \alpha_2, a, Q \right)$ and $N_{I}\left(\alpha_1, \alpha_2, a, Q \right)$ using Equations \eqref{EqNINL} and \eqref{normalizeF}. Note that $\bar{\nu}_{X}<\bar{\nu}_{I}<\bar{\nu}_{M}$, where $\bar{\nu}_{X} < \bar{\nu}\left(r \right) < \bar{\nu}_{I}$ outside ISCO ($r_I < r < r_X$) and $\bar{\nu}_{I}< \bar{\nu}\left(r \right) < \bar{\nu}_{M}$ inside ISCO ($r_M < r < r_I$).

\begin{figure}[H]
\centering
\mbox{ \hspace{-0.5cm}
 \subfigure[]{
\includegraphics[scale=0.45]{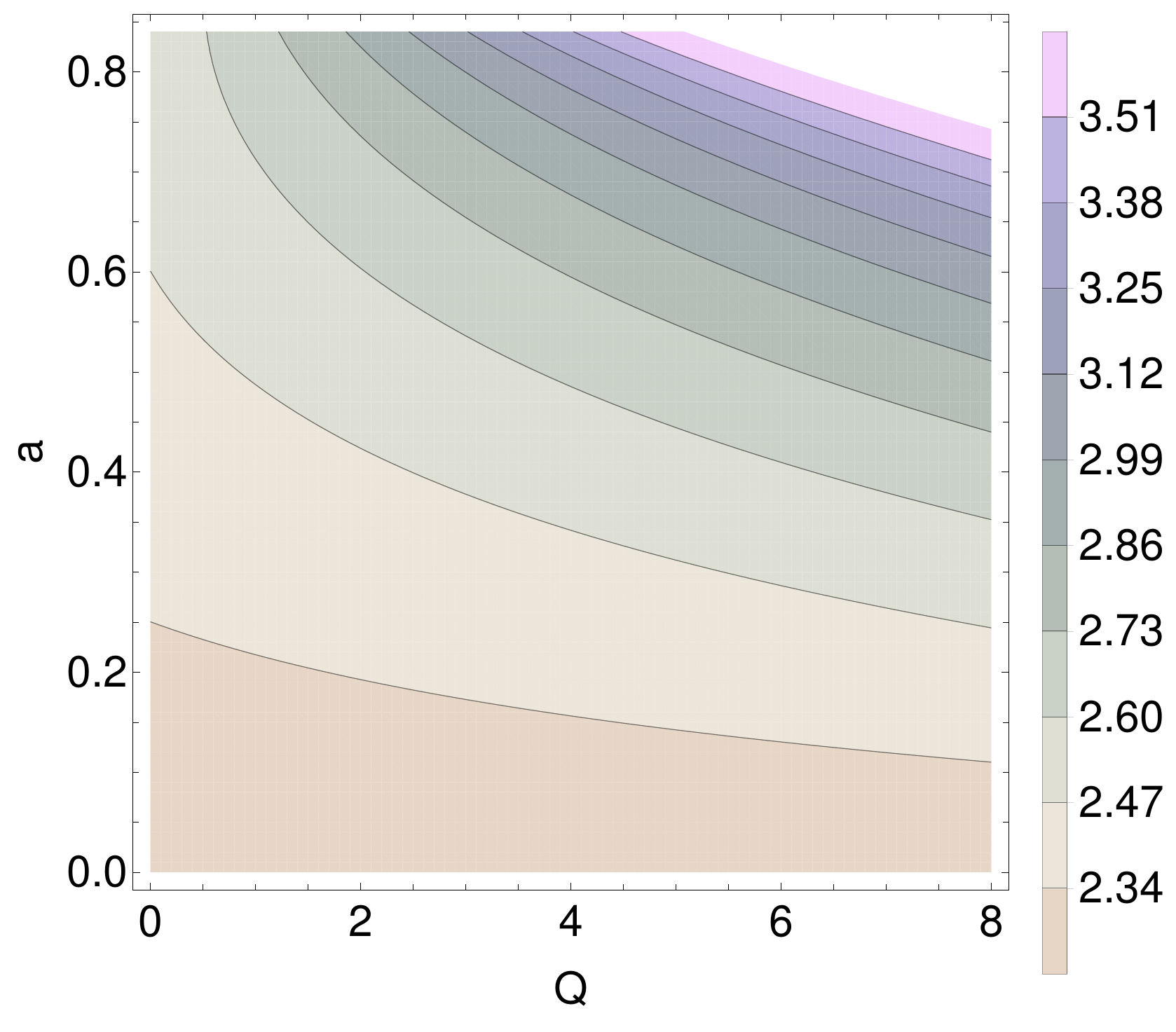}}
\hspace{0.7cm}
\subfigure[]{
\includegraphics[scale=0.45]{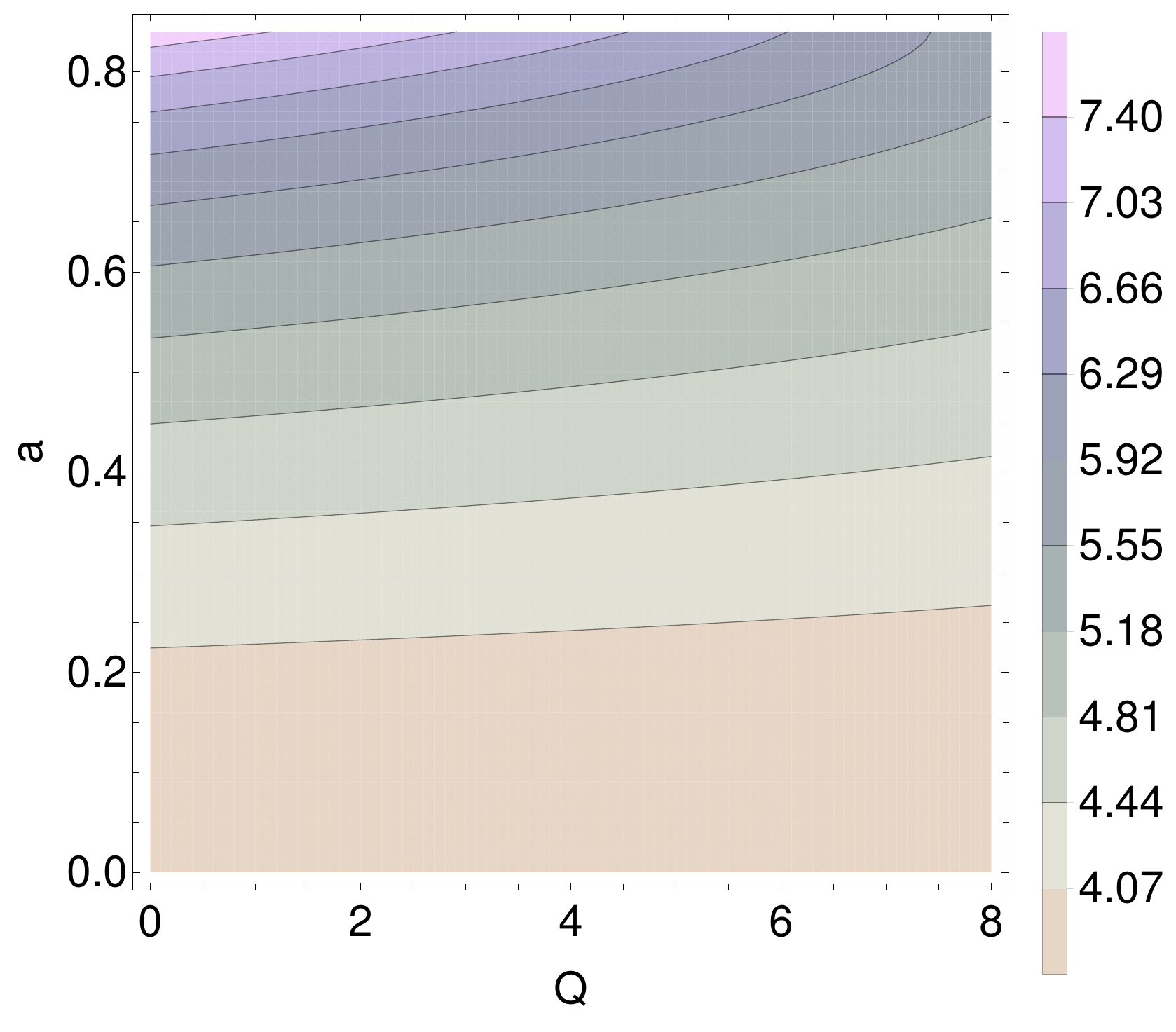}}}
\caption{\label{alphaplots}The figure shows contours of (\textbf{a}) $\alpha_1$, and (\textbf{b}) $\alpha_2$ in the ($Q$, $a$) plane.}
\end{figure}

\begin{figure}[H]
\centering
\mbox{
 \subfigure[]{
 \hspace{-0.4cm}
\includegraphics[scale=0.38]{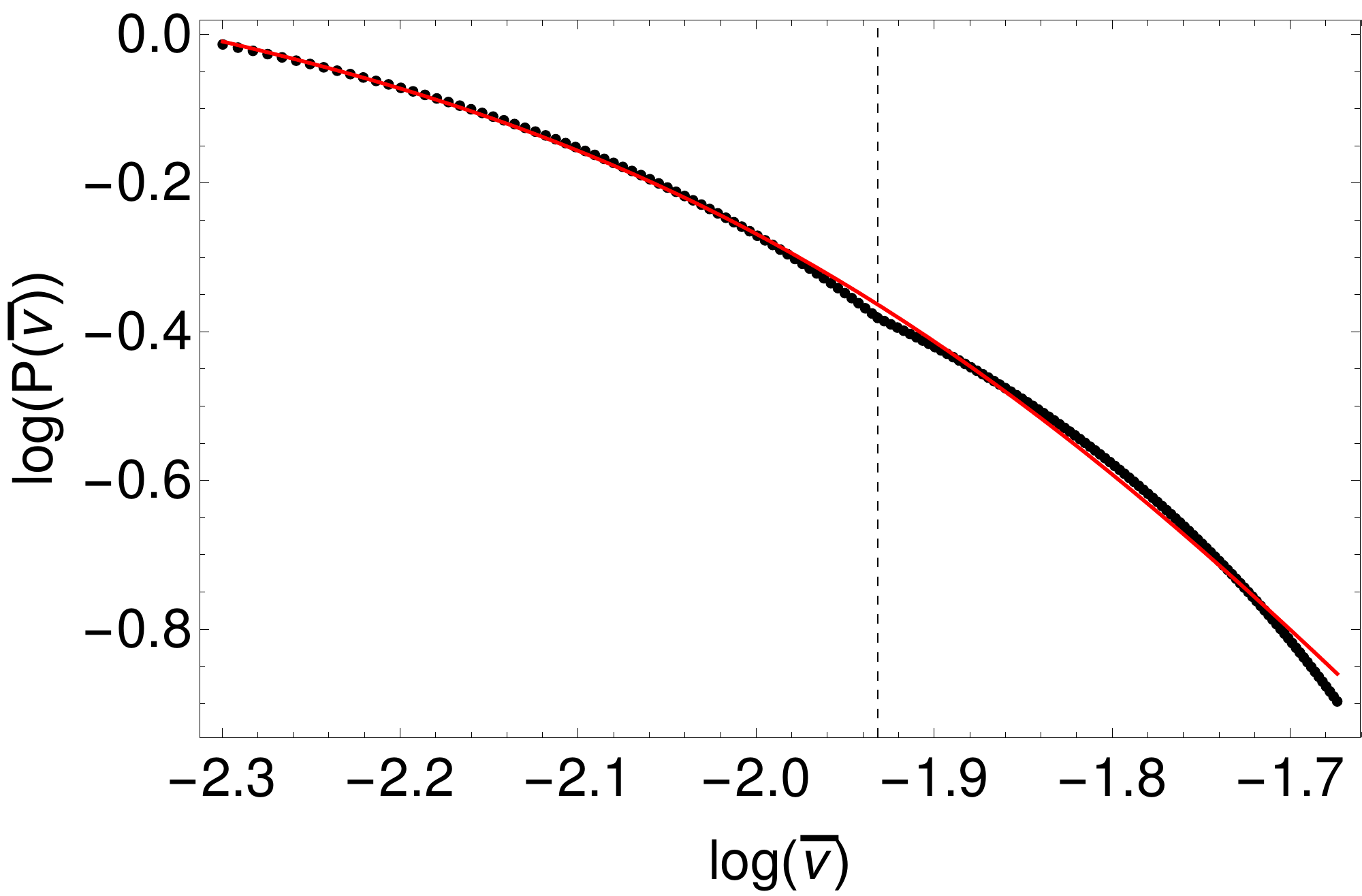}}
\hspace{0.3cm}
\subfigure[]{
\includegraphics[scale=0.38]{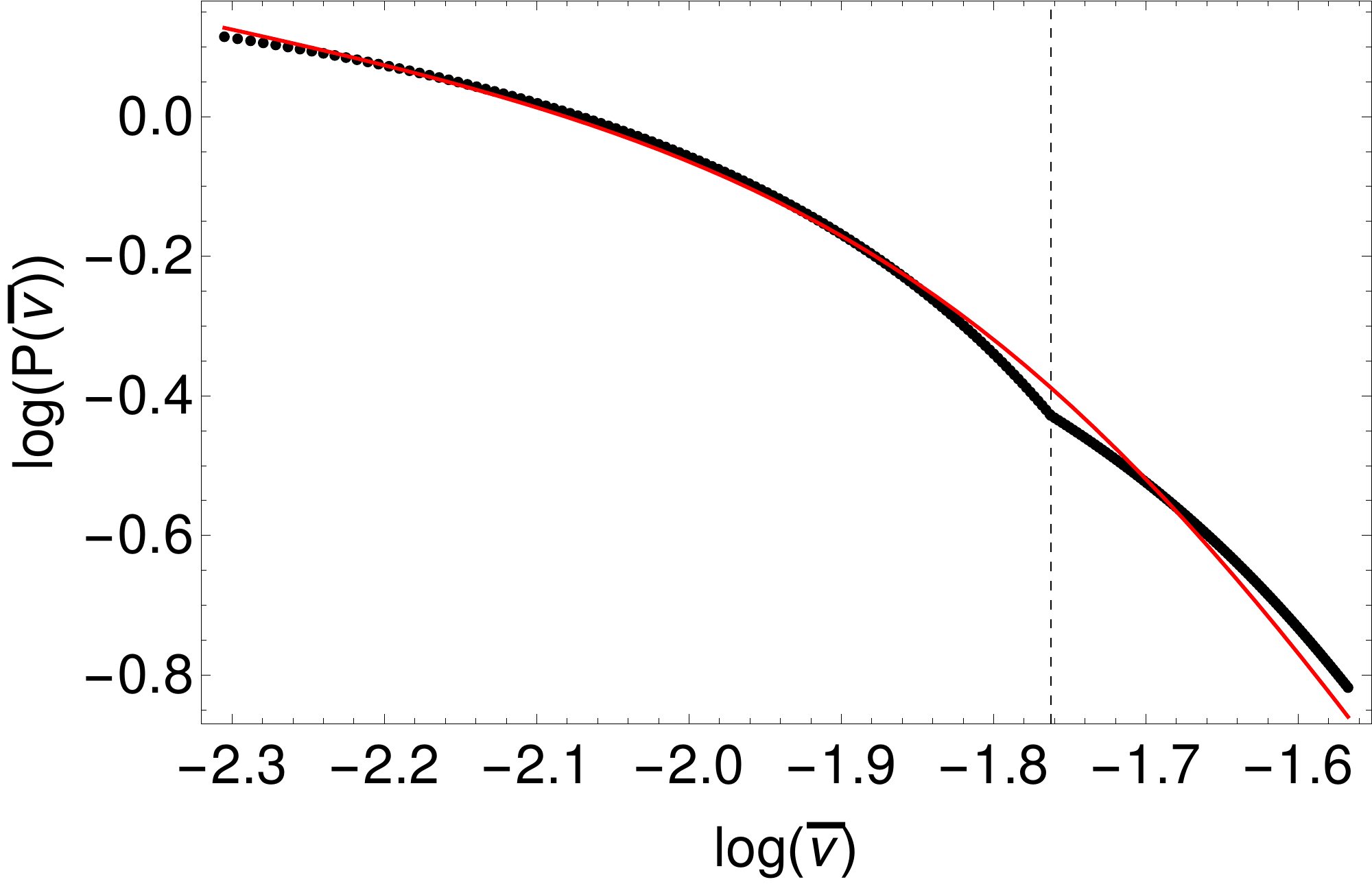}}}
\
\caption{\label{PSDplot}The figure shows examples of PSD, $P(\bar{\nu}) \propto F(\bar{\nu})^2$, profile obtained using the ROM for the parameter combinations (\textbf{a}) $\#$1, and (\textbf{b}) $\#$2  given in Table \ref{alpha12tab}. The red curve shows the bending power-law model fit, given by Equation \eqref{bendlaw}, where the fitting parameters are shown in Table \ref{alpha12tab}. The vertical black dashed curve corresponds to the ISCO (break) frequency.}
\end{figure}

Now, we describe the procedure to obtain the model parameters $\alpha_1$ and $\alpha_2$ using observations:
\begin{enumerate}
\item If $\beta_1$ is the average slope of the observed PSD after the break frequency, $\bar{\nu}>\bar{\nu}_{b}$, given by
\begin{subequations}
\begin{equation}
\frac{\Delta \log[P\left( \bar{\nu}\right)]}{\Delta \log[\bar{\nu}]}=\beta_1 \ \ \ \Rightarrow \ \ \ 2\frac{\Delta \log[F_1\left( \bar{\nu}\right)]}{\Delta \log[\bar{\nu}]}=\beta_1,
\end{equation}
where $\Delta$ represents the difference of values at the end points defined by MBSO and ISCO in our model: the end point of the PSD for $\bar{\nu}>\bar{\nu}_{b}$ (where $\bar{\nu}_{b}=\bar{\nu}_{I}\left(a  \right)$) is at the MBSO radius ($\bar{\nu}_{M} \left(a \right)$), so that
\begin{equation}
2 \log\left[\frac{f_1\left(\alpha_1, \alpha_2,  \bar{\nu}_{M} \left(a\right) , a, Q\right)}{f_1\left( \alpha_1, \alpha_2, \bar{\nu}_{I} \left(a \right), a, Q\right)} \right] =\beta_1 \log \left[\frac{\bar{\nu}_{M} \left(a \right)}{\bar{\nu}_{I} \left(a \right)}  \right],
\end{equation}
\begin{equation}
f_1\left( \alpha_1, \alpha_2, \bar{\nu}_{M} \left(a  \right), a, Q\right)= u_1 \left( a, \beta_1\right), 
\end{equation}
\begin{equation}
\Rightarrow \left[ 1-\frac{C_{1IM}\left(\alpha_1, a, Q\right)V\left(a, \alpha_1, \alpha_2 \right)}{N_I\left(\alpha_1, \alpha_2, a, Q \right)}\right]=u_1 \left( a, \beta_1\right),
\end{equation}
where 
\begin{equation}
u_1 \left( a, \beta_1\right)=\left(\frac{\bar{\nu}_{M} \left(a\right)}{\bar{\nu}_{I} \left(a\right)}\right)^{\beta_1/2},
\end{equation}
where $N_{I}\left(\alpha_1, \alpha_2,  a, Q\right)$ can be substituted using Equation \eqref{EqNINL}, which yields
\begin{equation}
\frac{1- u_1 \left( a, \beta_1\right)}{ 1+ u_1 \left( a, \beta_1\right)} =  V\left(a, \alpha_1, \alpha_2 \right) C_{1IM}\left(\alpha_1, a, Q\right). \label{relation1}
\end{equation}
\end{subequations}
Hence, for a given combination of \{$a$, $Q$\}, we obtain a relation, given by Equation \eqref{relation1}, where \{$\alpha_1$, $\alpha_2$\} are unknowns.
\item Similarly, if $\beta_2$ is the average slope of the observed PSD before the break frequency, $\bar{\nu}<\bar{\nu}_b$, we have
\begin{subequations}
\begin{equation}
2\frac{\Delta \log[F_2\left( \bar{\nu}\right)]}{\Delta \log[\bar{\nu}]}=\beta_2.
\end{equation}
The lower extreme of the PSD at $r=r_X$, for $\bar{\nu}<\bar{\nu}_b$, is given by $\bar{\nu}_{X}\left(a  \right)$, so that
\begin{equation}
2 \log \left[\frac{f_2\left(  \alpha_1, \alpha_2, \bar{\nu}_{I} \left(a\right) , a, Q\right)}{f_2\left(  \alpha_1, \alpha_2, \bar{\nu}_{X} \left(a\right), a, Q\right)} \right] =\beta_2 \log \left[\frac{\bar{\nu}_{I} \left(a\right)}{\bar{\nu}_{X} \left(a \right)}  \right],
\end{equation}
\begin{equation}
n_X\left( \alpha_1, \alpha_2, a, Q\right)= u_2 \left( a, \beta_2\right), 
\end{equation}
where
\begin{equation}
u_2 \left( a, \beta_2\right)=\left(\frac{\bar{\nu}_{I} \left(a \right)}{\bar{\nu}_{X} \left(a\right)}\right)^{-\beta_2/2}.
\end{equation}
The substitution of $N_{I}\left(\alpha_1, \alpha_2,  a, Q\right)$ and $N_{X}\left(\alpha_1, \alpha_2,  a, Q\right)$ using Equations \eqref{EqNINL} and \eqref{normalizeF} gives
\begin{equation}
\frac{u_2 \left( a, \beta_2\right) -1}{ 1+ u_1 \left( a, \beta_1\right)} =  W\left(a, \alpha_1, \alpha_2 \right) C_{2XI}\left(\alpha_2, a\right). \label{relation2}
\end{equation}
\end{subequations}
which is another relation to solve for \{$\alpha_1$, $\alpha_2$\}. Hence, Equations \eqref{Nnu1}, \eqref{relation1}, and \eqref{relation2} together give us values for \{$\alpha_1$, $\alpha_2$\} for a fixed combination of \{$a$, $Q$\}.
\item We compute the slopes \{$\alpha_1$, $\alpha_2$\} by the above mentioned criteria for different combinations of ($a$, $Q$), which are shown in Table \ref{alpha12tab}. We find that $\alpha_1$ ranges between $\sim$[$2.3-4$] and $\alpha_2$ is in the range $\sim$[$3.7-8.9$], indicating that a power-law model for the intrinsic mechanical energy of the orbiting matter describes the shape of the observed PSD reasonably well. Additionally, if we reverse the analysis to estimate \{$\beta_1$, $\beta_2$\} by fixing \{$\alpha_1=2.5$, $\alpha_2=3.5$\} for ($a=0.5$, $Q=2$), we find \{$\beta_1=-1.97$, $\beta_2=-0.77$\} which are in good agreement with observations. We also show contours of $\alpha_1$ and $\alpha_2$ in the ($Q$, $a$) plane in Figure \ref{alphaplots}, where the values of $\alpha_1$ and $\alpha_2$ increase with $a$. We also see that contours are independent of $Q$ for small $a$, which is expected because the non-equatorial orbits do not exist in Schwarzschild spacetime.
\item The examples of PSD profile obtained in the scaled frequency space, $\bar{\nu}$, are shown in Figure \ref{PSDplot}. We~see that the PSD profiles for given parameter combinations in Table \ref{alpha12tab} show good fits to the expected bending power-law model, Equation \eqref{bendlaw}. The PSD represents a general power spectrum obtained independent of the mass of the black hole; hence, it applies to the stellar-mass black holes also. This~validates the ROM as a plausible model for PSD observed in black holes. 
\end{enumerate}
\vspace{-10pt}

\begin{table}[H]
\caption{The table summarizes the computed values of ($\alpha_1$, $\alpha_2$) and the parameter fits to the bending power-law, Equation \eqref{bendlaw}, for various combinations of ($a$, $Q$), where we fixed $r_X=10$ and \{$\beta_1=-2$, $\beta_2=-1$\}, and the frequencies were scaled by $(c^3/GM_{\bullet})$.}
\centering
 \tablesize{\normalsize}
\begin{tabular}{ccccccc}
\toprule
\textbf{\#}&\textbf{($a$, $Q$)} & \textbf{$\alpha_1$}  & \textbf{$\alpha_2$} & \textbf{$\alpha_l$} & \textbf{$\alpha_h$} & \textbf{$P_0$}\\
\midrule
1 &($0.1, 0$)& $2.286$ & $3.753$ & 0.282 & 2.74 & 0.866\\
2& ($0.5, 4$)& $2.615$ & $4.864$ & 0.413 & 3.453 & 0.818\\
3& ($0.9, 0$)& $2.462$ & $8.873$ & 0.488 & 5.561 & 1.112\\
4& ($0.9, 4$)& $3.944$ & $7.407$ & 0.497 & 5.328 & 0.925\\
\bottomrule
\end{tabular}
\label{alpha12tab}
\end{table}

\section{Summary}
\label{summary}
The results are summarized below:
\begin{itemize}
\item In Section \ref{intro}, we summarized the observations for X-ray QPOs, which are traditionally associated with the accretion disk and corona, $\gamma$-ray QPOs normally attributed to a jet, and the X-ray PSD usually connected with the inner and outer corona (see Figure \ref{modelplot}).
\item In Section \ref{XrayQPOmodel}, we motivated the creation of (G)RPM models for X-ray QPOs and extracted the spins and radii for the sources, listed in Table \ref{AGNXrayQPO}, based on the model given in \cite{Stella1999a,Stella1999b,RMQPO2020}. The GRPM model confirms that the detected QPO in Type-2 AGN 2XMM J123103.2+110648 is an LFQPO, as it was also suggested by \cite{Linetal2013}. In a statistical analysis, we were able to determine these parameters and their errors for 1H 0707-495, the case of two simultaneous QPOs, based on the observed QPO frequencies and their errors. The~results are presented in Table \ref{1H07circtable} for circular orbits and in Table \ref{1H07sphtable1} for spherical orbits. We found non-planar orbits, with $Q\sim(1-12)$, which are very close to a Kerr black hole, that ($r_s\sim(8.2-8.3)$; $a\sim0.14$) are the possible solutions for QPO frequencies of 1H 0707-495.
\item Next, in Section \ref{Jetmodel}, we applied the relativistic kinematic jet model to check its validity by comparing the basic frequency with the observed QPO periods in BL Lac objects, given in Table \ref{AGNjetQPO}. The ratio $T_0/T_F$ is typically in the range $1-20$, which is reasonable, given the range of footpoint radii of the field lines and typical location of the Alfv\'{e}n point up to which the field line is rigid \cite{MohanMangalam2015}. It motivates detailed relativistic MHD models along with polarization profile predictions (as given in \cite{Mangalam2018}) to compare with observations.
\item In Section \ref{PSDmodel}, we built a relativistic orbit model consisting of circular and spherical orbits that have a power-law distribution, and its mechanical energy is split into two parts (above and below the energy at ISCO). This formulation leads to unique results relating to the PSD slopes (before ($\bar{\nu}<\bar{\nu}_{b}$) and after ($\bar{\nu}>\bar{\nu}_{b}$) the break) with those of the energy spectrum for the given spin and mass of the black hole (Figures \ref{alphaplots} and \ref{PSDplot}). We plan to test this model against observations to extract \{$a$, $M_{\bullet}$\}.
\end{itemize}
\section{Discussion and Conclusions}
\label{Discussion}
We add the following points of discussion of our results and conclusion:
\begin{enumerate}
\item The periastron and nodal precession of the particle orbits is an intrinsic phenomenon in Kerr geometry, which is a consequence of strong gravity and axisymmetry of the spacetime. We propose in the GRPM \cite{RMQPO2020} that the precession frequencies of matter blobs orbiting in these trajectories, very close to the Kerr black hole, modulate the X-ray flux, from the thin accretion disk where the flow is hot. The origin of these non-equatorial orbits of blobs in a slim torus region having a single radius is motivated in \cite{RMQPO2020}, where a model of fluid flow in the general relativistic thin accretion disk \cite{Pennaetal2012} is studied. In this study, we suggest that the edge region, defined in \cite{Pennaetal2012}, is a launchpad for plasma instabilities, where blobs orbit with fundamental frequencies of the geodesics near the edge and in the geodesic region (defined in \cite{Pennaetal2012}), in which Hamiltonian dynamics is applicable. We also show in the GRPM that these geodesics span a torus region, which overlaps with the edge and geodesic region of \cite{Pennaetal2012}.  
\item The QPOs in NLSy1s are usually observed when $L/L_{Edd}$ is very high; for example, $L/L_{Edd} \sim10$ in the case of RE J1034+396 \cite{Gierlinski2008} implies a high accretion rate, but the association of $L/L_{Edd}$ with the QPO frequencies is not clear. Moreover, even if one assumes that the accretion process in AGN and BHXRB is the same and that both show similar characteristic $\mathcal{Q}$ shape in the hardness-intensity diagram \cite{Remillard2006}, over a timescale, $T$, this would be $10^5$$-$$10^6$ times more than BHXRB timescales, as $T \propto M_{\bullet}$.
\item Our relativistic orbit model (ROM) is built on the formulation of the intrinsic mechanical energy distribution of the plasma in motion, where three frequencies $\nu_{X}<\nu_{I}<\nu_{M}$ correspond to the low-frequency end, break frequency, and the high-frequency end of the PSD. However, there is a noise component to be added at higher frequencies of the PSD to obtain a more realistic PSD shape to the intrinsic energy distribution related to the frequencies of the unstable orbits inside MBSO. A more generalized approach will be to incorporate frequencies of the more general eccentric and non-planar orbits ($e\neq0$, $Q\neq0$) contributing to the PSD shape. This is planned as future work.
\item The fundamental frequencies of the spherical geodesics in the Kerr geometry seem to explain the PSD in the Inner Corona (IC) region, where $P\left( \nu_{I}< \nu < \nu_{M}\right)$; whereas the frequencies of the Outer Corona (OC) region are associated with the circular orbits, where $P\left( \nu_{X}< \nu < \nu_{I}\right)$. The results of this toy statistical model, ROM, seem promising. A detailed physical model is required to predict the power law indices in the energy spectrum. Furthermore, including a more ellaborate transfer function taking into account the GR effects like light bending and Doppler boosting, is in order for further study.
\item The paradigm of the ROM can be tested against observations by extracting \{$M_{\bullet}$, $a$\} from observed \{$\nu_{X}, \nu_{I},\nu_{M}, \beta_1, \beta_2$\}, and by exploring the parameter space \{$\alpha_1$, $\alpha_2$\} which is the basis of the PSD for the ROM model. In the future, we plan to apply and test this model against several observed PSD of various AGN sources.
\item The total power of a PSD having a power-law profile is given by
\begin{eqnarray}
\mathcal{P}_T \propto  \int^{\nu_c}_{0} \left(\frac{\nu}{\nu_c}\right)^\tau  {\rm d} \nu \  \propto \nu_c  \int^{1}_{0} X^{\tau} {\rm d} X \ \propto  \nu_c , 
\end{eqnarray}
where $X=\nu / \nu_c$, $\tau$ is the power-law index, and $\nu_c$ is the upper frequency cut-off of the PSD. On the other hand, from the Wiener--Khinchin theorem, $\mathcal{P}_T=F_{var}^2 \propto \left(\sigma^2 - \sigma_N^2\right)$, where $F_{var}$ gives a measure of the time signal variance above the noise and $\sigma_N^2$ is the variance in the noise measurable from observations. This gives the relation between the measured quantity and $\nu_c$ as, $F_{var} \propto \nu_c^{0.5}$, where the cutoff $\nu_c$ provides a measure of the spin and mass of the black hole if the disk cuts off at the ISCO or MBSO radius; this implies $\nu_c^{0.5} \propto M_{\bullet}^{0.5}$. Using a more complicated PSD distribution expected from the ROM and using $F_{var}$, we can give better estimates for $\nu_c$ and hence extract \{$M_{\bullet}$, $a$\}, using $F_{var}$, and study statistical trends from a sample of sources with known \{$M_{\bullet}$, $a$\}. 
\end{enumerate}
%%%%%%%%%%%%%%%%%%%%%%%%%%%%%%%%%%%%%%%%%%
%%%%%%%%%%%%%%%%%%%%%%%%%%%%%%%%%%%%%%%%%%
%%%%%%%%%%%%%%%%%%%%%%%%%%%%%%%%%%%%%%%%%%
%\section{Patents}
%This section is not mandatory, but may be added if there are patents resulting from the work reported in this manuscript.

%%%%%%%%%%%%%%%%%%%%%%%%%%%%%%%%%%%%%%%%%%
\vspace{6pt} 
%%%%%%%%%%%%%%%%%%%%%%%%%%%%%%%%%%%%%%%%%%
\authorcontributions{\textbf{Prerna Rana}: Conceptualization, Methodology, Investigation, Software, Writing - original draft. \textbf{A. Mangalam}: Conceptualization, Methodology, Investigation, Writing - original draft, review and editing, Supervision. All authors have read and agreed to the published version of the manuscript.}%
%%%%%%%%%%%%%%%%%%%%%%%%%%%%%%%%%%%%%%%%%%
\funding{We acknowledge DST SERB CRG grant number 2018/003415 for financial support.}
%%%%%%%%%%%%%%%%%%%%%%%%%%%%%%%%%%%%%%%%%%
\acknowledgments{We would like to thank the anonymous referees for detailed and insightful suggestions that have improved our paper significantly. We would like to thank Saikat Das for helping us with Figure \ref{modelplot}. We also thank Alok Gupta and Paul Wiita for useful discussions. We acknowledge the use and support of the IIA-HPC facility.}
%%%%%%%%%%%%%%%%%%%%%%%%%%%%%%%%%%%%%%%%%%
\conflictsofinterest{The authors declare no conflict of interest.}
%%%%%%%%%%%%%%%%%%%%%%%%%%%%%%%%%%%%%%%%%%
\abbreviations{The following abbreviations are used in the manuscript:\\
\noindent 
\begin{tabular}{@{}ll}
AGN & Active Galactic Nuclei \\
BHXRB & Black Hole X-ray Binaries \\
ULX & Ultra-Luminous X-ray source \\
QPO & Quasi-Periodic Oscillation \\
IC & Inner Corona \\
OC & Outer Corona \\
ISCO & Innermost Stable Circular Orbit \\
MBCO & Marginally Bound Circular Orbit \\
MBSO & Marginally Bound Spherical Orbit \\
NLSy1 & Narrow-Line Seyfert 1 \\
GRPM & General Relativistic Precession Model\\
ROM & Relativistic Orbit Model \\
PSD & Power Spectral Density \\
\end{tabular}}

\reftitle{References}
\end{document}